\documentclass[12pt,a4paper,fleqn]{article}
\usepackage[utf8]{inputenc}
\usepackage{authblk}
\usepackage{geometry}
\usepackage{xcolor}
\usepackage{natbib}
\usepackage{graphicx}
\usepackage{pdflscape}
\usepackage{appendix}
\usepackage{setspace}
\usepackage{makecell}
\usepackage{graphbox}

\usepackage{colortbl}

\usepackage{threeparttable}
\usepackage{booktabs}

\usepackage{amssymb}
\usepackage{amsbsy}
\usepackage{bm}
\usepackage{amsmath}
\usepackage{amsthm}
\usepackage{graphicx}
\usepackage{mathtools}

\usepackage{enumitem}
\newlist{steps}{enumerate}{1}
\setlist[steps, 1]{label = Step \arabic*:}

\usepackage{dcolumn}
\newcolumntype{d}[1]{D{.}{.}{#1}}



\usepackage{caption}
\usepackage{subcaption}
\definecolor{nblue}{HTML}{000660}
\usepackage[colorlinks=true,urlcolor=nblue,linkcolor=nblue,citecolor=nblue]{hyperref}
\addtocounter{page}{-1}

\title{\LARGE \textbf{Model Specification for Bayesian Neural Networks in Macroeconomics}}
\title{\LARGE \textbf{Bayesian Neural Learning of Nonlinearities in Macroeconomics and Finance}}
\title{\LARGE \textbf{Bayesian Neural Networks for Macroeconomic Analysis}\thanks{
\noindent Corresponding author: Massimiliano Marcellino. Department of Economics, Bocconi University. \textit{Address}: Via Roentgen 1, 20136 Milano, Italy. \textit{Email}: \href{mailto:massimiliano.marcellino@unibocconi.it}{massimiliano.marcellino@unibocconi.it}. The authors thank the Editors Serena Ng and George Kapetanios, three anonymous referees, Jamie Cross, Martin Feldkircher, Laurent Ferrara, Christian Hotz-Behofsits, Malte Knueppel, Michael McCracken, Carlos Montes-Galdon, Michael Pfarrhofer, and participants at the CFE 2022 conference, COMPSTAT 2022 conference, 12$^{th}$ ECB Conference on Forecasting Techniques 2023, and IIF MacroFor 2023 webinar for helpful comments and suggestions.  Hauzenberger gratefully acknowledges financial support from the Jubiläumsfonds of the Oesterreichische Nationalbank (OeNB, grant no. 18718, 18763, and 18765), and Huber acknowledges financial support from the Austrian Science Fund (FWF, grant no. ZK 35) and the Jubiläumsfonds of the OeNB (grant no. 18304). The views expressed in this paper do not necessarily reflect those of the Oesterreichische Nationalbank or the Eurosystem.}}
\author[1]{\MakeUppercase{Niko Hauzenberger}}
\author[2]{\MakeUppercase{Florian Huber}}
\author[3]{\MakeUppercase{Karin Klieber}}
\author[4]{\MakeUppercase{Massimiliano Marcellino}}
\affil[1]{\textit{University of Strathclyde}}
\affil[2]{\textit{University of Salzburg}}
\affil[3]{\textit{Oesterreichische Nationalbank}}
\affil[4]{\textit{Bocconi University, IGIER, CEPR, Baffi-Carefin and BIDSA}}

\begin{document}

\maketitle\thispagestyle{empty}\normalsize\vspace*{-2em}\small

\begin{center}
\begin{minipage}{0.8\textwidth}
\noindent\small  Macroeconomic data is characterized by a limited number of observations (small $T$), many time series (big $K$) but also by featuring temporal dependence. Neural networks, by contrast, are designed for datasets with millions of observations and  covariates. In this paper, we develop Bayesian neural networks (BNNs) that are well-suited for handling datasets commonly used for macroeconomic analysis in policy institutions. Our approach avoids extensive specification searches through a novel mixture specification for the activation function that appropriately selects the form of nonlinearities. Shrinkage priors are used to prune the network and force irrelevant neurons to zero. To cope with heteroskedasticity, the BNN is augmented with a stochastic volatility model for the error term. We illustrate how the model can be used in a policy institution by first showing that our different BNNs produce precise density forecasts, typically better than those from other machine learning methods. Finally, we showcase how our model can be used to recover nonlinearities in the reaction of macroeconomic aggregates to financial shocks.\\ \\ 

\end{minipage}
\end{center}

\spacing{1.5}\normalsize\renewcommand{\thepage}{\arabic{page}}

\newpage

\section{Introduction}
Policy making in central banks and other governmental institutions is often informed by economic models that are, for simplicity, assumed to be linear or take relatively simple nonlinear forms. For instance, the Phillips curve is a key analytical framework for the analysis and conduct of monetary policy. This relationship, however, is often assumed to be linear and empirical specifications have been shown to forecast poorly \citep[see, e.g.][]{clark2006predictive}. The weak out-of-sample predictive power is often attributed to structural breaks in the coefficients or other forms of nonlinearities and researchers increasingly adopt nonlinear parametric  models to estimate Phillips curves  \citep[see, e.g., ][]{benigno2023s}.  Another example is the empirical evidence on the effects of financial shocks on the economy. Linear models often estimate a transitory disruptive effect of financial shocks \citep{gilchrist2012credit, boivin2020dynamic} on output. However, several recent papers \citep{barnichon2022effects, mumtaz2022impulse} find that adverse financial disruptions are much more harmful than benign disruptions. This is because linear models mix over shocks of both signs and thus under-/over-estimate the actual effect of an adverse/a benign financial shocks. 


These  examples highlight that evidence-based policy making could benefit from more flexible models. The main obstacle to using such flexible models is that they often require strong assumptions on the nature of nonlinearities and this requires substantial prior information. The question,  however, is whether a particular form of nonlinearity is really appropriate or whether it implies model mis-specification. Deciding on the appropriate form of nonlinearities is thus crucial for accurate inference, and modern machine learning techniques can be applied to learn the functional form from the data.

In this paper, we develop Bayesian neural networks (BNNs) for macroeconomic policy analysis. NNs have the advantage of being able to approximate any form of nonlinear conditional mean relation arbitrarily well \citep{hornik1989neuralnet} and, in a wide range of different fields \citep[see, e.g.,][]{kourentzes2013intermittent,wen2017multi,salinas2020deepar,sezer2020nnfin}  have been shown to be particularly useful for forecasting purposes. Moreover, they nest other popular methods in machine learning such as trees \citep[see][]{wang2020uncertainty}. Their use in econometrics, however, has been limited, some recent relevant examples include \cite{farrell2021deep, chronopoulos2023forecasting, chronopoulos2023deep}. This is due to several reasons, and we aim to address (at least) three  of these through our model.

First, NNs are trained on vast datasets and often possess millions of free parameters. For instance, the MNIST dataset \citep{lecun1998gradient} includes $T=60,000$ observations and $K=784$ covariates. By contrast, macroeconomic datasets such as the popular \cite{mccracken2016fred} database feature a few hundred observations and almost as many possible time series. Hence, specifying and estimating fully-fledged multi-layer NNs on such datasets is challenging, which makes their use for empirical analyses and policy making problematic.  For instance, applications require to specify the number of hidden layers, the number of neurons,  the form of the activation function, the algorithm as well as associated nuisance parameters used for training the models.  

To address these issues, we aim to minimize the possible range of competing choices by introducing stochastic model selection techniques to decide both the type of (layer specific) activation functions and the number of neurons per layer. The former is achieved through a novel mixture model that averages over a set of pre-specified activation functions. The vast majority of NN papers assume instead a single activation function per layer, and a common choice with good theoretical properties is the Rectified Linear Unit \citep[ReLU, see][]{polson2018posterior, farrell2021deep}. Yet, this choice might be restrictive, and our mixture specification provides additional flexibility, without the need to carry out cross-validation or repeated estimation of the model. Moreover, we select the number of neurons by using Bayesian shrinkage priors that force individual neurons aggressively to zero.  In particular, we follow \cite{ghosh2019model, bhadra2020horseshoe} and use a horseshoe prior which does not depend on additional tuning parameters. As a result, our proposed BNN only requires the researcher to decide on the number of hidden layers.

Second, to speed-up computations, NNs are typically estimated  using maximum a-posteriori (MAP), minimum mean squared error (MMSE), or variational Bayes (VB) inference.  However, uncertainty quantification with these techniques is difficult and often relies on approximations. {For instance, VB only approximates the joint posterior distribution of the model and neglects any parameter uncertainty in the predictive distributions, resulting in underestimating the actual predictive variance. Moreover, capturing departures from non-Gaussianity (such as heavy tails, multi-modality, or skewness) would also be challenging.} To cope with this, we opt for a fully Bayesian estimation approach that samples from the exact full conditional posterior distributions.\footnote{Exact conditional on the MCMC algorithm to have converged to the correct stationary distribution.} In our model, objects of interest, such as predictive distributions, can be then obtained through Monte Carlo integration. The resulting predictive densities can be highly non-Gaussian and exhibit features such as heavy tails, multiple modes or skewness. 

Third, NNs are often viewed as a black box, with limited scope for structural economic analysis. To show that this is not necessarily the case, we develop BNN based local projections \citep[LPs, see][]{jorda2005estimation}.  These LPs allow us to investigate how macroeconomic aggregates react to economic shocks in a nonlinear way, and to tease out implied dynamics and asymmetries in the responses.  

After developing theoretical tools for the specification, estimation, forecasting and structural analysis for BNNs, we assess their empirical performance, first with synthetic data, and then with actual US macro data. To generate synthetic data that closely resemble actual macroeconomic dynamics, we use a data generating process (DGP) inspired by the  nonlinear Phillips curve in \cite{benigno2023s}. It turns out that shallow models with flexible activation functions recover the mean function of inflation with more precision than deep models. 

Next, we show that our BNNs produce accurate short-term density forecasts for key US economic variables (inflation, industrial production and employment), often better than those from standard models commonly used in econometrics (such as dynamic models augmented with principal components), and machine learning (such as LASSO, elastic net, (Bayesian) additive regression trees, and random forests). In addition, for all the variables we consider, deep BNNs produce much better in-sample fit than shallow BNNs, but differences in density forecast accuracy are very small, which supports the use of the latter which are computationally simpler. Moreover, our mixture activation function yields generally only modest gains, but it frees the researcher from the necessity to decide on one particular activation function and thus reduces the number of inputs to the model. Furthermore, a recursive evaluation suggests that nonlinearities pay off during turbulent times, such as the recession in the early 2000s, the global financial crisis (GFC) and the onset of the Covid-19 pandemic. 

Finally, we use our nonlinear machine learning models to investigate how key US macroeconomic quantities react to different economic shocks, and assess whether the reactions are non-proportional with respect to the size of the shock and asymmetric with respect to its sign. Specifically, we develop BNN-based local projections (LPs) and investigate how financial shocks --- measured with the excessive bond premium \citep[EBP,][]{gilchrist2012credit} --- impact US inflation, industrial production and employment. We find substantial asymmetries with respect to sign, with negative shocks exerting stronger effects on inflation and employment than positive ones, but substantial proportionality with respect to the size of the shock.

The remainder of the paper is structured as follows. Section \ref{sec:econometrics} develops our BNN model for use with macroeconomic data. Section \ref{sec: synthetic} illustrates the model using synthetic data that resembles US inflation dynamics. Section \ref{sec: emp_work} considers forecasting US economic variables. Section \ref{sec: structural analysis} analyzes the propagation of financial shocks via NN based LPs. The last section summarizes and concludes the paper. An Online Appendix provides additional technical details.

\section{Bayesian neural networks for macroeconomic data}\label{sec:econometrics}
This section develops our general Bayesian neural network model. After discussing key model specification issues in Sub-section \ref{sec: likelihood}, we consider approximations in Sub-section \ref{sec:approximation}, choice of the activation function in Sub-section \ref{sec:activation}, modeling the error variance in Sub-section \ref{sec: SV}, suitable Bayesian regularization priors in Sub-section \ref{sec: prior}, and posterior computation in Sub-section \ref{sec: posterior}.

\subsection{A general nonparametric regression model} \label{sec: likelihood}
Our goal is to estimate an unknown and possibly nonlinear relationship between a macroeconomic  time series $y_t \in \mathbb{R}$ and $K$ covariates $\bm x_t \in \mathbb{R}^K$. In what follows, the covariates in $\bm x_t$ can include instruments for structural economic shocks, lags of $y_t$ or other (lagged) macroeconomic and financial indicators.  In our setting, $K$ can be very large relative to the sample size $T$ and the relationship between $y_t$ and $\bm x_t$ possibly highly nonlinear.  

We assume a general nonlinear regression given by:
\begin{equation}
y_t = \bm x_t'\bm \gamma + f(\bm x_t) + \varepsilon_t, \quad \varepsilon_t \sim \mathcal{N}(0, \sigma_t^2),\label{eq: reg_general}
\end{equation}
where $\bm \gamma$ is a vector of $K$ linear coefficients, $f: \mathbb{R}^K \to \mathbb R$ is a function of unknown (nonlinear) form and $\varepsilon_t$ is a Gaussian shock with zero mean and time-varying variance $\sigma_t^2$. A typical assumption is that $f$ is $\alpha-$H\"{o}lder smooth on $[0, 1]^K$.\footnote{The assumption that $\bm x_t$ is restricted to the $K$-dimensional unit hypercube is not restrictive in practice since we can appropriately transform any covariate to fulfill this.} The class of $\alpha-$H\"{o}lder smooth functions is defined by $\mathcal{H}^\alpha_K = \{f: [0, 1]^K \to \mathbb{R}; ||f||_\mathbb{H}^\alpha < \infty\}$ where $||f||_\mathcal{H}^\alpha$ denotes the Hölder norm \citep{Schmidt-Hieber, polson2018posterior}. The parameter $\alpha > 0$ indicates the smoothness of the class of functions we aim to approximate.  

The inclusion of a linear part in the model is meant to use the nonlinear component only to capture proper nonlinear relationships between the target and the explanatory variables. Recent evidence suggests that linear models are competitive in terms of forecast accuracy. It is mostly during turbulent times (such as recessions) that adding nonlinear components pays off \citep[e.g.,][]{clark2023tail}. We take this empirical fact into account through the linear piece which we expect to explain the vast majority of variation over the estimation period.

Similarly, the presence of a time-varying error variance reduces the risks that nonlinearities in the conditional mean show up simply to capture outliers or periods of high volatility. It also implies that our model adapts to situations not learned through the conditional mean by increasing $\sigma_t^2$, thus inflating uncertainty surrounding predictions to provide a proper assessment of the point forecasts reliability, which matters particularly when the latter are used for policy making. The assumption of Gaussian shocks is not restrictive when combined with stochastic volatility (SV), but it would be feasible to incorporate more flexible error distributions based on scale-location mixtures of Gaussians \citep[see, e.g., ][]{escobar1995bayesian}.

In macroeconomics and finance,  $f$ is often assumed to be known. For instance, if $f(\bm x_t) = 0$ for all $t$ we end up with a constant parameter regression model. Another commonly used model arises  if $f(\bm x_t) = \bm x_t' \bm \gamma_t$ with $\bm \gamma_t$ denoting $K$ time-varying parameters (TVPs). Other specifications which can be seen as special cases of Eq. (\ref{eq: reg_general}) are threshold and Markov switching models \citep[see, e.g., ][]{Hamilton1989switching, tong1990nonlinear, terasvirta1994transition}, polynomial regression \citep[see, e.g., ][]{mccrary2008discontinuity, lee2010regression} or models with interaction effects \citep[see, e.g., ][]{ai2003interaction, imbens2009interaction, greene2010testing}. 

This brief discussion shows that the choice of $f$ is one of the most important modeling decisions the researcher needs to take. In this paper, we follow a different route and estimate $f$. This can be achieved through nonparametric techniques such as Bayesian additive regression trees \citep[see e.g.,][]{chipman2010bart, huber2020nowcasting, clark2023tail, clark2024investigating}, random forests \citep[see, e.g.,][]{coulombe2020randomforest}, Gaussian processes \citep[see e.g., ][]{williams2006gaussian, crawford2019approx, hauzenberger2021gaussian}, splines \citep[see, e.g., ][]{vasicek1982term, engle2008spline} or wavelets \citep[see, e.g., ][]{ramsey1998wavelet, gallegati2008wavelet}. 

\subsection{Deep neural network approximation}\label{sec:approximation}
In this paper, we  aim to learn $f$ using a NN. In many areas outside economics, NNs have been used with some success \citep[see ][ for surveys on time series forecasting with deep learning]{lara2021experimental,lim2021time}. In this paper, our goal is to design neural networks that can be used in combination with macroeconomic data and its associated particularities. These relate to the fact that the data is often persistent, features structural breaks and volatility clustering but also that the length of the time series is short.  Hence, researchers and practitioners alike can not rely on theoretical results in the literature \citep[see, e.g.,][]{polson2018posterior, Schmidt-Hieber, farrell2021deep} that mostly assume $T$ to be huge but also rule out autocorrelation in the data. Hence, the large-scale NNs that are typically used are often not suited for macro datasets, and this motivates the main modeling choices we take.

We use a deep NN to approximate the function $f$. This approximating model consists of $L$ layers, with each layer featuring $Q_\ell~(\ell=1, \dots, L)$ neurons. The nodes are related through an activation function $h_{\ell, q}~(q=1, \dots, Q_\ell)$. In what follows, we let $\bm h_\ell = (h_{\ell, 1}, \dots, h_{\ell, Q_\ell})'$ denote an activation function that can be layer and neuron-specific.\footnote{In the discussion we do not add bias terms to simplify notation. In our empirical work, all networks include a bias term as well.} In this case, our approximation to $f$ can be written as:
\begin{equation}
    f(\bm x_t) \approx \widehat{f}_L(\bm x_t) = \bm W_{L+1} \bm h_L\left( \bm W_L \bm h_{L-1} (\cdots \bm W_2 \bm h_1 (\bm W_1 \bm x_t))\right),\label{eq: DNN}
\end{equation}
with $\bm W_\ell$ denoting a $Q_\ell \times Q_{\ell-1}$ matrix of network weights for $\ell=2,\dots,L$, while $\bm W_{L+1}$ is a $1 \times Q_L$ vector and $\bm W_1$ is a $Q_1 \times K$ matrix. In the subsequent discussion, we set $Q_\ell = Q$ for $\ell=1 ,\dots, L$.
It is worth noting that if $L=1$, we end up with a shallow neural network:
\begin{equation*}
    \widehat{f}_1(\bm x_t) = \bm W_2 \bm h_1(\bm W_1 \bm x_t).
\end{equation*}

Both shallow and deep neural networks serve as universal approximators. However, as outlined in \cite{mhaskar2017and}, deep NNs attain a certain level of approximation error with exponentially fewer parameters than  shallow NNs. If the true function $f$ is compositional, meaning that $f(\bm x_t) = h_J(h_{J-1}(\dots h_1(\bm x_t)))$, a shallow neural network would require a very large number of neurons to approximate it with an arbitrary approximation error $\epsilon>0$.  For the macro series we are interested in, it is, however, not clear whether the true underlying structural model implies a compositional structure and, in addition, given that the informational content in typical datasets is relatively low compared to the vast dimensional datasets used in other fields, we will treat the shallow NN as  a serious competitor in the analysis that follows. 

Typically, researchers treat the weights as the only unknown parameters in the network and set $L$ and $\bm h_l$ a priori and based on best practice or cross-validation. The choice of the activation function could have important consequences for the empirical performance of the model \citep{karlik2011performance, agostinelli2014learning} and, in our general specification, the activation function is not only layer but also neuron-specific. Hence, the space of possible combinations of activation functions for a moderate number of hidden layers becomes intractable. A practicable solution is to set $h_{\ell, q} = h$ for all $\ell$ and $q$. In the literature, a standard choice with excellent theoretical properties  is the rectified unit activation function $\text{ReLU}(x) = \text{max}(0, x)$.  However, theoretical results in, e.g., \cite{polson2018posterior} or \cite{farrell2021deep} are based on situations where $T \to \infty$ and without assuming dependence in the responses. For macroeconomic data that are often analyzed in actual policy contexts, these conditions are not fulfilled and the small sample properties of the ReLU activation functions are not well understood. As a solution, one contribution of this paper is to treat $h_{\ell, q}$ as a discrete hyperparameter which we estimate alongside the weights $\bm W_\ell$ and other model parameters.  We turn to this issue in the next sub-section.

\subsection{Convex combinations of activation functions}\label{sec:activation}
Instead of using a single activation function for the whole network, we construct a mixture specification that averages over four commonly used activation functions:  leakyReLU (1), sigmoid (2), ReLU (3) and hyperbolic tangent (tanh, 4). We let $h^{(m)}$ with $m \in \{\text{leakyReLU}, \text{sigmoid}, \text{ReLU}, \text{tanh}\}$ denote an activation function out of this set of different functions. Each specific choice $h^{(m)}$  has different implications on the flexibility of the neural network to capture nonlinearities in the data. Table \ref{tab:act_funcs} provides a summary of the functions used.  

\begin{table}[htb!]
\caption{Set of activation functions.}\label{tab:act_funcs}
{\scriptsize
\begin{center}
\begin{tabular*}{\textwidth}{ll @{\extracolsep{\fill}} cc}
\toprule
& \bfseries Activation function  & \bfseries Equation & \bfseries Plot \tabularnewline
\midrule
(1) & LeakyReLU & 
    $h^{(1)}(x) = \begin{cases}
        0.01x & x < 0 \\
        x & x \geq 0
        \end{cases}$ & \includegraphics[align=c,scale=.3]{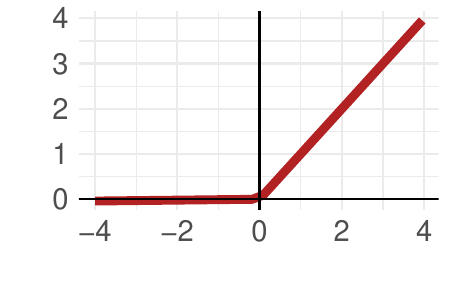} \tabularnewline
\midrule
(2) & Sigmoid &  $h^{(2)}(x) = \frac{1}{1+\exp{(-x)}}$  & \includegraphics[align=c,scale=.3]{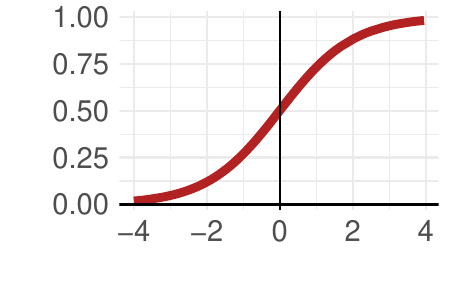} \tabularnewline
\midrule
(3) & Rectified Linear Unit (ReLU) & $h^{(3)}(x) = \max{(0,x)}$ & \includegraphics[align=c,scale=.3]{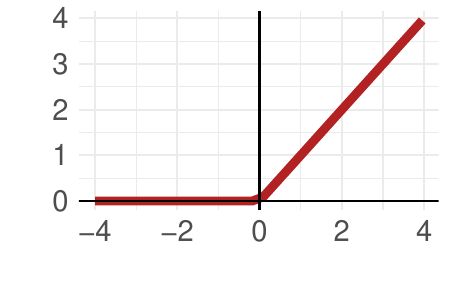} \tabularnewline
\midrule
(4) & Hyperbolic tangent (tanh) & $h^{(4)}(x) = \frac{\exp{(x)}-\exp{(-x)}}{\exp{(x)}+\exp{(-x)}}$ & \includegraphics[align=c,scale=.3]{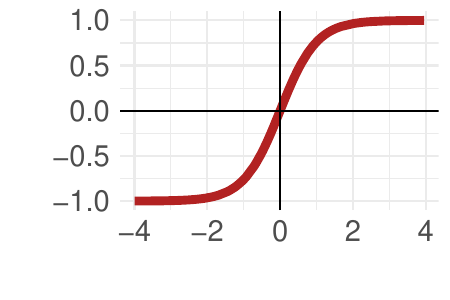} \tabularnewline
\bottomrule
\end{tabular*}
\end{center}}
\end{table}

We assume that each activation function is given by:
\begin{equation}
        h_{\ell, q}(z_{\ell q, t}) = \sum_{m=1}^4  {\omega}^{(m)}_{\ell, q} h^{(m)}(z_{\ell q, t}),
\end{equation}
with $z_{\ell q, t}$ being the $q^{th}$ element in the recursively defined vector $\bm z_{\ell, t} = \bm W_{\ell} \bm h_{\ell-1}( \bm z_{\ell-1, t})$ and $\bm z_{1, t} = \bm W_1 \bm x_t$. The neuron and layer-specific weights ${\omega}^{(m)}_{\ell, q}$ satisfy ${\omega}^{(m)}_{\ell, q} \ge 0$ and $\sum_m {\omega}^{(m)}_{\ell, q} = 1$.  The main implication of this specification is that we do not need to decide on one particular activation function but we average over a set of prominent activation functions. The weights reflect the relative importance of a specific activation function over other options.  In principle, if a researcher has prior knowledge that a given activation function should be nonlinear (either through observing features of the data or theoretical knowledge), one can use this information through a suitable prior on the weights ${\omega}^{(m)}_{\ell, q}$. 

An alternative way of writing this mixture specification is through a discrete  random variable $\delta_{\ell, q} \in \{1, \dots, 4\}$. The probability that $\delta_{\ell, q} = m$ is then set equal to:
\begin{equation*}
    \text{Prob}(\delta_{\ell, q} = m) = {\omega}^{(m)}_{\ell, q}.
\end{equation*}
Using mixtures of activation functions increases the flexibility substantially without introducing a large number of additional parameters. Our approach is related to evolutionary neural networks \citep{yao1999evolving, turner2014neuroevolution} that learn the network structure adaptively but, as we will discuss below, allows for fully-fledged uncertainty quantification on the activation functions and the weights. 

\begin{figure}[t!]
\caption{Nonlinearities in the nexus between inflation and money growth. \label{fig:nonlinearities_actfunc_shlw}}
\centering
\begin{minipage}{0.32\textwidth}
\centering
\scriptsize \textit{linear}
\end{minipage}
\begin{minipage}{0.32\textwidth}
\centering
\scriptsize \textit{sigmoid}
\end{minipage}
\begin{minipage}{0.32\textwidth}
\centering
\scriptsize \textit{tanh}
\end{minipage}

\begin{minipage}{0.32\textwidth}
\centering
\includegraphics[scale=.4]{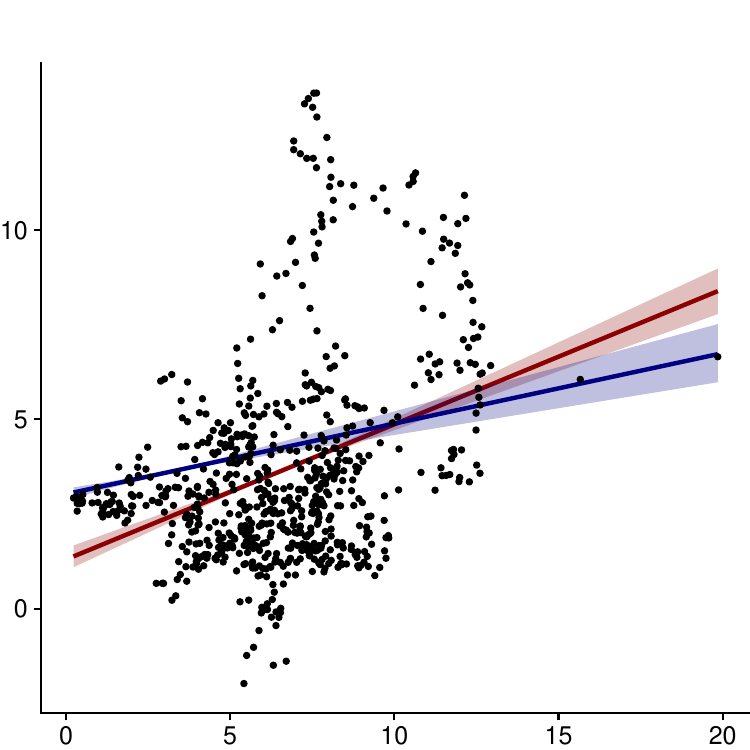}
\end{minipage}
\begin{minipage}{0.32\textwidth}
\centering
\includegraphics[scale=.4]{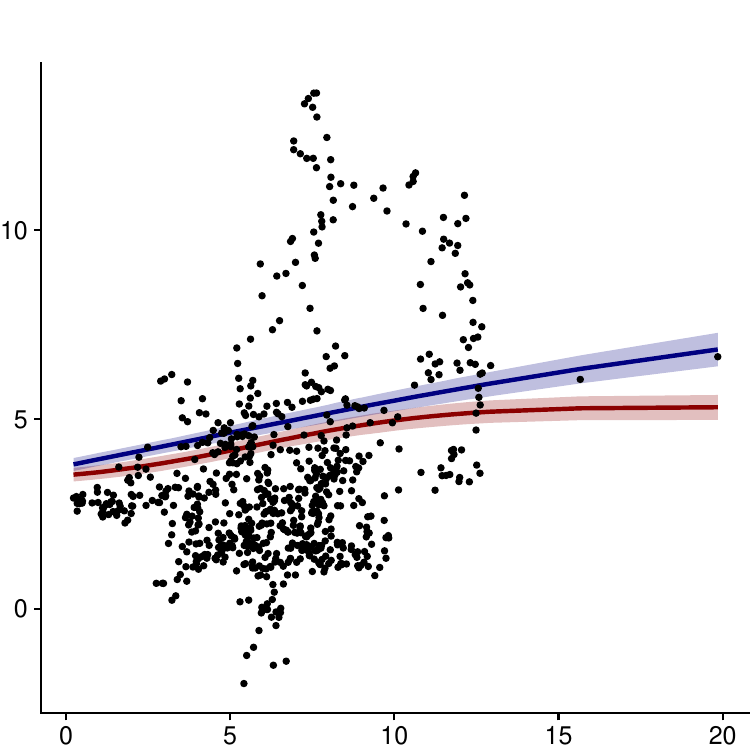}
\end{minipage}
\begin{minipage}{0.32\textwidth}
\centering
\includegraphics[scale=.4]{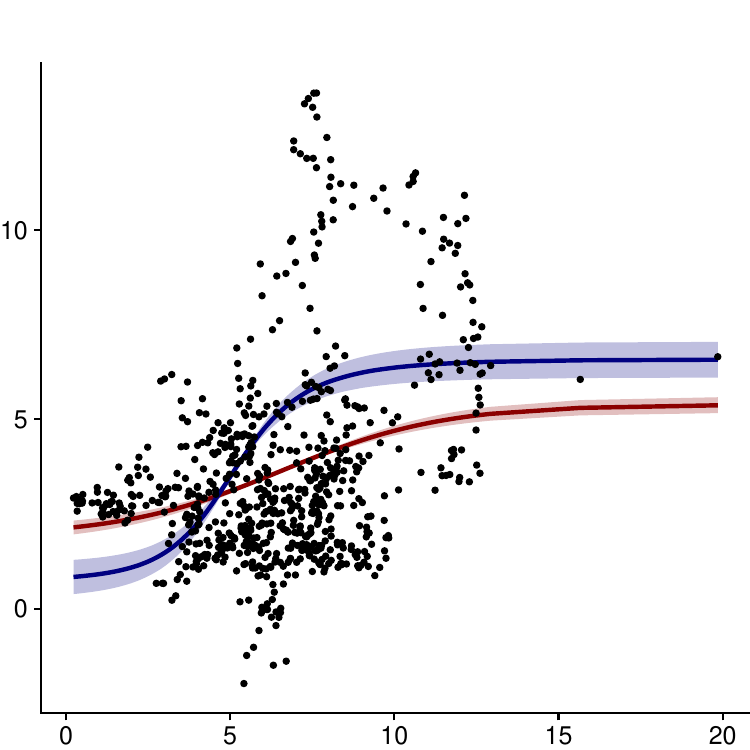}
\end{minipage}

\vspace{20pt}
\begin{minipage}{0.32\textwidth}
\centering
\scriptsize \textit{ReLU}
\end{minipage}
\begin{minipage}{0.32\textwidth}
\centering
\scriptsize \textit{leakyReLU}
\end{minipage}
\begin{minipage}{0.32\textwidth}
\centering
\scriptsize \textbf{\textit{convex combination}}
\end{minipage}

\begin{minipage}{0.32\textwidth}
\centering
\includegraphics[scale=.4]{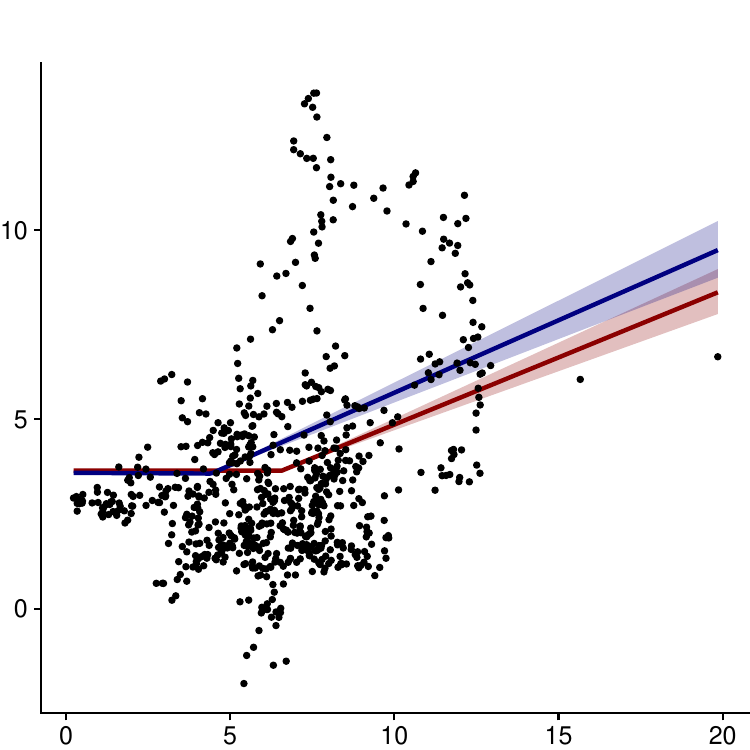}
\end{minipage}
\begin{minipage}{0.32\textwidth}
\centering
\includegraphics[scale=.4]{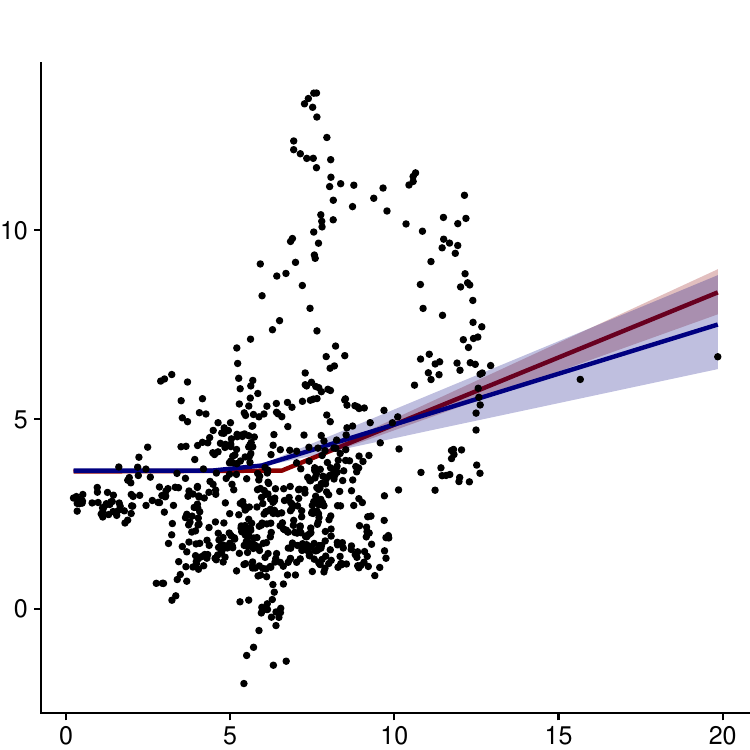}
\end{minipage}
\begin{minipage}{0.32\textwidth}
\centering
\includegraphics[scale=.4]{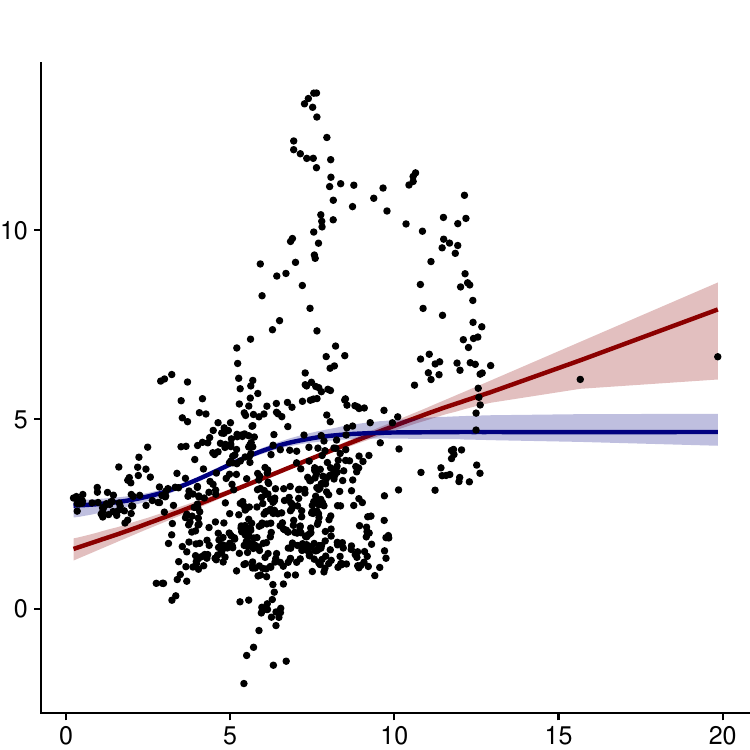}
\end{minipage}

\begin{minipage}{0.32\textwidth}
\centering
\vspace{-20pt}
\hspace*{-33pt}\includegraphics[scale=.4]{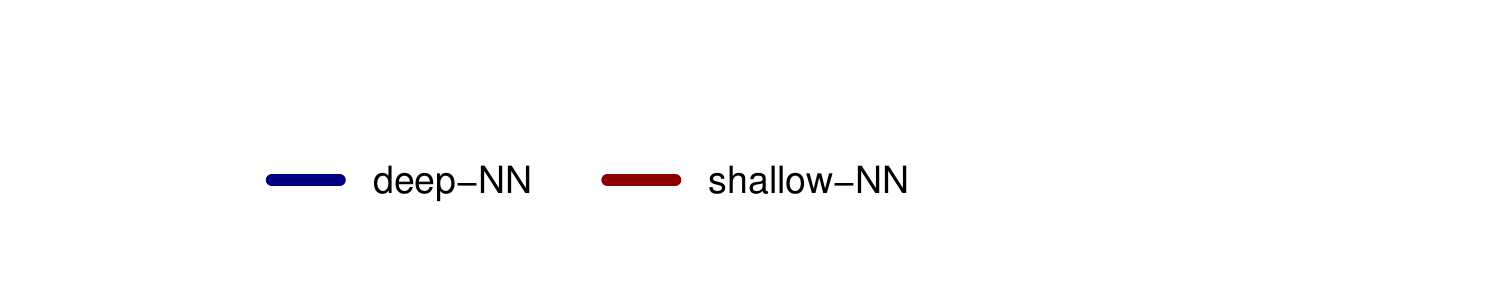}
\end{minipage}

\vspace{-10pt}
\begin{minipage}{\textwidth}
\scriptsize
\emph{Note:} This figure shows the nexus between inflation and money growth for the US and illustrates the functional forms of the activation functions specified in Table  \ref{tab:act_funcs}. The data for the consumer price index (i.e., CPIAUCSL) and money supply (i.e., M2SL) are taken from the FRED-MD database as described in \cite{mccracken2016fred}. We plot the following example: $y_t = f(x_{t}) + \varepsilon_t$, where $y_t$ is the year-on-year inflation and $x_{t}$ is the $18^{th}$ lag of the year-on-year money supply growth (see Eq. (\ref{eq:illustr})). $f$ refers to the activation functions specified in Table \ref{tab:act_funcs}. The bottom-right panel `convex combination' refers to our main specification, where we let the data decide on the form of the activation function. 
\end{minipage}
\end{figure}

Depending on the choice of the weights, we can get combined activation functions that share features of each of the four individual activation functions. We illustrate the effect different activation functions have on the function estimates using a simple univariate example. This example models the relationship between US year-on-year inflation ($y_t$) and the year-on-year money growth rate ($x_t$) in a nonlinear manner. These two series are obtained from the FRED-MD database \citep{mccracken2016fred}. To account for the leading effect of money growth on inflation, we specify $x_t$ as the $18^{th}$ lag of money growth \citep[see, e.g., ][]{reichlin2007short, amisano2013money}. The corresponding nonparametric regression is given by:
\begin{equation}\label{eq:illustr}
    y_t = f(x_t) + \varepsilon_t, \quad \varepsilon_t \sim \mathcal{N}(0, \sigma^2).
\end{equation}
We compare the effect that different activation functions have on the mean estimate $f(x_t)$ in  Figure \ref{fig:nonlinearities_actfunc_shlw} by considering two BNNs estimated using the techniques outlined in the next sub-section. The first is a shallow one that sets $L=1$ and includes only a single neuron. The red shaded areas represent the $5^{th}$ and $95^{th}$ percentiles of the posterior of $\widehat{f}_1(x_t)$ while the red line is the posterior median. The second model is a deep BNN with $L=3$ and a single neuron for all $\ell$ as well. The corresponding posterior percentiles are depicted in blue. 
In this figure, the (lagged) values of money growth are on the x-axis while the yearly change in inflation is on the y-axis. Considering the linear specification (i.e., $f(x_t) = \beta x_t)$ suggests a positive relationship between money growth and inflation. Nonlinear activation functions, both for the deep and shallow BNN, often yield estimates of mean relations that display substantial nonlinearities. This finding holds for tanh, ReLU and leakyReLU. There are cases where the deep model produces more nonlinearities (such as for tanh) or where thresholds change (in the case of ReLU) but the shape is often similar and suggests that for low levels of money growth, inflation displays only modest reactions. Notice that for leakyReLU, both the deep and shallow BNN produce similar mean relations, indicated by overlapping posterior credible intervals. 

The differences between shallow and deep models increase if we consider our convex mixture activation function. In this case, the shallow neural network produces a fit close to the linear model. Notice, however, that for large levels of money growth, the uncertainty surrounding the functional estimates increases appreciably. By contrast, the deep BNN with flexible activation functions produces an estimate of the conditional mean that implies increasing levels of inflation for money growth between three and eight percent. For values larger than eight percent, the relationship flattens out and the model predicts no substantial fluctuations of inflation. 

This short, stylized example illustrates that the different activation functions give rise to different, albeit similar, estimated mean relationships. Since in all our empirical work we set $Q$ to a large value and use a large panel of covariates, the models we propose are capable of extracting complex nonlinear features in a very flexible manner.
\subsection{Adding stochastic volatility} \label{sec: SV}
Neural networks often explicitly or implicitly assume that the error variance is constant. This assumption implies that the mean function explains a constant share of variation in $y_t$ over time. For macroeconomic data, this assumption is strong. In exceptional periods such as the GFC or the Covid-19 pandemic, not only mean relations can change but larger shocks than usual can hit the economy. Hence, we model the error variance in a time-varying manner using a standard stochastic volatility model.

Our model assumes that $\nu_t = \log \sigma_t^2$ evolves according to an AR(1) process:
\begin{equation}
    \nu_t = \mu_\nu + \rho_\nu (\nu_{t-1} - \mu_\nu) + \varsigma_t, \quad \varsigma_t \sim \mathcal{N}(0, \xi_\nu^2), \label{eq:sv}
\end{equation}
with $\mu_\nu$ denoting the long-run level of the log-volatility, $\rho_\nu$ the persistence parameter and $\xi_\nu^2$ the state equation variance. This model assumes that the error variance evolves rather smoothly over time and features its own shock. For later convenience, we let $\bm \nu = (\nu_1, \dots, \nu_T)'$ denote the full history of the log-volatilities and $\bm \beta_\nu = (\mu_\nu, \rho_\nu, \xi_\nu^2)'$ the parameters of the log-volatility state equation.

It is worth stressing that we opt for a SV specification because it is flexible and  has been shown to work well for macroeconomic data \citep{clark2011real}.  Alternative approaches such as the Generalized Conditional Heteroskedastic model \citep[GARCH,][]{bollerslev1986generalized} are also feasible. In principle, it would also be possible to use a NN to estimate the volatility process as well \citep[see, e.g.,][]{coulombe2023reactive,nguyen2023statistical}. 

\subsection{Achieving shrinkage in deep neural networks} \label{sec: prior}
We opt for a Bayesian approach to estimating the NN. This involves specifying suitable priors on the parameters of the model. The prior setup we use introduces regularization on the weights so that we can effectively use a large number of neurons per hidden layer but avoid overfitting issues by shrinking weights associated with irrelevant neurons to zero. 

On the weights of the network we consider a shrinkage prior. Early contributions propose shrinkage and regularization priors on regression models that imply nonlinear transformations of the covariates but linearity in the parameters \citep[see, e.g.][]{tipping2001sparse}. \cite{titterington2004bayesian} discusses shrinkage priors in NNs and comes up with a prior on the weights of the form:
\begin{equation*}
    p(\bm W_\ell|\phi_{{\ell, ij}}) \propto \exp \left\lbrace - \frac{1}{2} \sum_{i} \sum_j \phi^2_{\ell, i j} w_{\ell, i j} \right\rbrace,
\end{equation*}
with $\phi_{{\ell, ij}}$ denoting a shrinkage hyperparameter that controls the degree of shrinkage. Hence, large values of $\phi_{{\ell, ij}}$ implies that the  $(i, j)^{th}$ element of $\bm W_\ell, w_{\ell, i j}$, should be strongly forced to zero. Empirically, we often face a situation where most $\phi_{\ell, ij}'$s are large and thus many elements in $\bm W_\ell$ are shrunk to zero, reducing the effective dimension of the state space considerably \citep{titterington2004bayesian}.

To capture this, global local shrinkage (GL) priors \citep{polson2010shrink} can be employed. These priors include a global shrinkage component that applies to groups of parameters and local components that allow for deviations in case of strong global shrinkage. A particular, hyperparameter-free, version of such a GL prior is the horseshoe prior \citep{carvalho2009handling}. The horseshoe has been first applied to NNs in  \cite{ghosh2019model} and \cite{bhadra2020horseshoe}. This prior is used to flexibly prune NNs without requiring prior information or introducing additional hyperparameters.  

Let $\bm w_{\ell, i \bullet}$ denote the $i^{th}$ row of $\bm W_\ell~(\ell=1, \dots, L+1)$. Both, on the linear coefficients in $\bm \gamma$ and $\bm w_{\ell, i \bullet}$ we  use horseshoe priors \citep{carvalho2010horseshoe}. The horseshoe implies the following prior hierarchy on each element of $\bm w_{\ell, i \bullet} = (w_{\ell, i1}, \dots, w_{\ell, i Q_{\ell -1}})'$:
\begin{equation*}
    w_{\ell, ij} \sim \mathcal{N}(0, \phi_{{\ell, ij}}), \quad \phi_{{\ell, ij}} = \lambda^2_{{\ell, i}} \varphi^2_{\ell, ij}, \quad \lambda_{{\ell, i}} \sim \mathcal{C}^+(0,1), \quad \varphi_{{\ell, ij}} \sim \mathcal{C}^+(0,1),
\end{equation*}
with $\lambda^2_{{\ell, i}}$ being a global (neuron-specific) shrinkage parameter which forces all elements in $\bm w_{\ell, i \bullet}$ towards the origin, $ \varphi_{\ell, ij}$ is a local scaling parameter that allows for coefficient-specific deviations in light of strong global shrinkage (i.e., if $\lambda^2_{\ell, i} \approx 0$).  Hence, within each layer we use the horseshoe to shrink the weights associated with irrelevant neurons to zero, effectively reducing the number of neurons per layer in a flexible manner. 

It is worth stressing that there is a close relationship between using a shrinkage prior on the weights and dropout, another popular technique for NN regularization. \cite{nalisnick2019dropout} discuss the equivalence between shrinkage priors and dropout.

Next, we need to discuss the prior on the activation weights $\omega^{(m)}_{\ell, q}$. In this case, we specify the prior probability that $\text{Prob}(\delta_q = m) = 1/4$. This choice implies that each activation function is equally likely a prior. In principle, alternative choices would be possible so that more weight is put on functions such as the ReLU where there exists strong evidence that this choice works well empirically.   

Finally, the prior on the parameters driving the SV processes are set along the lines suggested in \cite{kastner2014ancillarity}. This amounts to eliciting a Gaussian prior on $\mu_\nu \sim \mathcal{N}(0, 10)$, a Beta prior on $(\rho_\nu+1)/2 \sim \mathcal{B}(25, 1.5)$ and a Gamma prior on $\xi_\nu^2 \sim \mathcal{G}(1/2, 1/(2 s_\nu))$ where $s_\nu = 0.01$. The prior on the unconditional mean is relatively uninformative whereas the prior on the (transformed) persistence parameter pushes the latent process towards a random walk. The Gamma prior on $\xi_\nu^2$ is equivalent to a Normal prior on $\pm \xi_\nu$ with variance $s_\nu$. The choice $s_\nu=0.01$ implies some shrinkage on the amount of heteroskedasticity in the shocks.

\subsection{Posterior simulation}\label{sec: posterior}
In general, posterior inference in BNNs is extremely challenging. These challenges arise from the fact that typical input datasets are huge dimensional in $T$ and $K$. In our case, given the limited length of the time series, estimating a BNN is possible and estimation of such a model can be done within an hour on a standard desktop computer. 

At a general level, we draw from the joint posterior distribution using an MCMC algorithm that consists of several blocks. The most challenging block is related to the network weights for layers $\ell = 1, \dots, L$. In this case, we use state-of-the-art Hamiltonian Monte Carlo (HMC) techniques. The remaining blocks of our algorithm are standard and we provide more details in Section \ref{sec: fullpost} of the Online Appendix. Here, we only provide an overview of the algorithm. Our sampler cycles between the following steps:

\begin{itemize}
\item Both the linear coefficients $\bm \gamma$ of dimension $K$ and the weights vector of the output layer $\bm W_{L+1}$ of dimension $Q_L$, associated with the neurons, are obtained jointly from a standard multivariate Gaussian posterior, see Eqs. (\ref{eq:postbeta}) and (\ref{eq:postcons}).

\item For shrinking the linear coefficients, we use the horseshoe prior \citep{carvalho2010horseshoe} and update the corresponding hyperparameters by sampling from inverse Gamma distributions using the auxiliary sampler proposed in \cite{makalic2015simple}, see Eqs. (\ref{eq:postHScons1}) to (\ref{eq:postHScons4}). 

\item Sampling from $p(\bm W_\ell| \bullet)$, for $\ell = 1, \dots, L$, is achieved through an HMC step, see Eqs. (\ref{eq:hmc}) to (\ref{eq:MHstep}).

\item The shrinkage hyperparameters of the horseshoe prior on $\bm W_\ell$ are obtained from simple inverse Gamma posteriors, see Eqs. (\ref{eq:postHS1}) to (\ref{eq:postHS4}).

\item The function $h_{\ell, q}$ is simulated by first introducing an indicator $\delta_{\ell, q}$ which takes integer values one to four and indicates the precise function chosen. This indicator is then simulated from a multinomial distribution, see Eq. (\ref{eq:postMxInd}), and the relevant functional form of the activation function is chosen.

\item Draws of $\bm v$ and $\bm \beta_v$ are simulated using the algorithm proposed in \cite{kastner2014ancillarity}. 
\end{itemize}
We iterate through our MCMC algorithm 20,000 times and discard the first 10,000 draws as burn-in. 

\section{Illustration using synthetic data} \label{sec: synthetic}
Before we deal with actual data, it is worthwhile to investigate the properties of the different BNNs in terms of how well they recover the mean function. To do so, we need to come up with a data generating process (DGP) that closely resembles actual macroeconomic dynamics. This is done by setting up a DGP that is inspired by the model in \cite{benigno2023s}, which assumes that inflation follows a nonlinear Phillips curve of the form:
\begin{align*}
    \pi_t &= \mu_\pi + \rho_\pi \pi_{t-1} + \lbrace\beta_\pi + \beta_{\pi_d} D_t(\vartheta_t \ge 1)\rbrace \log \vartheta_t + \rho_{\upsilon} \upsilon_t + \sigma_{\pi} \epsilon_{\pi, t},  \\
    \log \vartheta_t &= \mu_\vartheta + \rho_{\vartheta, 1} \log \vartheta_{t-1} + \rho_{\vartheta, 2} \log \vartheta_{t-2} +\sigma_{\vartheta}\epsilon_{\vartheta, t}, \\
    \upsilon_t &= \sigma_{\upsilon} \epsilon_{\upsilon,t}, \\
    \epsilon_{j, t} &\sim \mathcal{N}(0, 1), \quad \text{ for } j \in \{\pi, \vartheta, \upsilon\},
\end{align*}
where $ \vartheta_t $ is a measure of labor market tightness and $D_t$ is a dummy variable taking value one if $\vartheta_t \ge 1 $. We take most parameters from the full-sample estimates of Table 1 in \cite{benigno2023s}. This implies setting $\mu_\pi = 0.192, \rho_\pi = 0.262, \beta_\pi = 0.222, \beta_{\pi_d}=3.896, \rho_{\upsilon}=0.0469$. $\sigma_\pi$ is set equal to $0.1$.   For the second equation, which  controls the dynamic evolution of the measure of labor market tightness, we set the parameters to closely match the dynamics of actual labor market tightness as proposed in \cite{NBERw30211} and used by \cite{benigno2023s}. This requires to set $\mu_\vartheta=-0.693, \rho_{\vartheta, 1}= 1.3, \rho_{\vartheta, 2}=-0.3$ and $\sigma_\vartheta = 0.05$. The third equation reflects a series of supply shocks ($\upsilon_t$) with $\sigma_{\upsilon} = 1$. From this DGP, we simulate $20$  time series with length $T=350$.

\begin{figure}[t!]
\caption{A single realization from the \cite{benigno2023s} DGP. \label{fig:pc_dgp}}
    \centering
    \includegraphics[scale=.7]{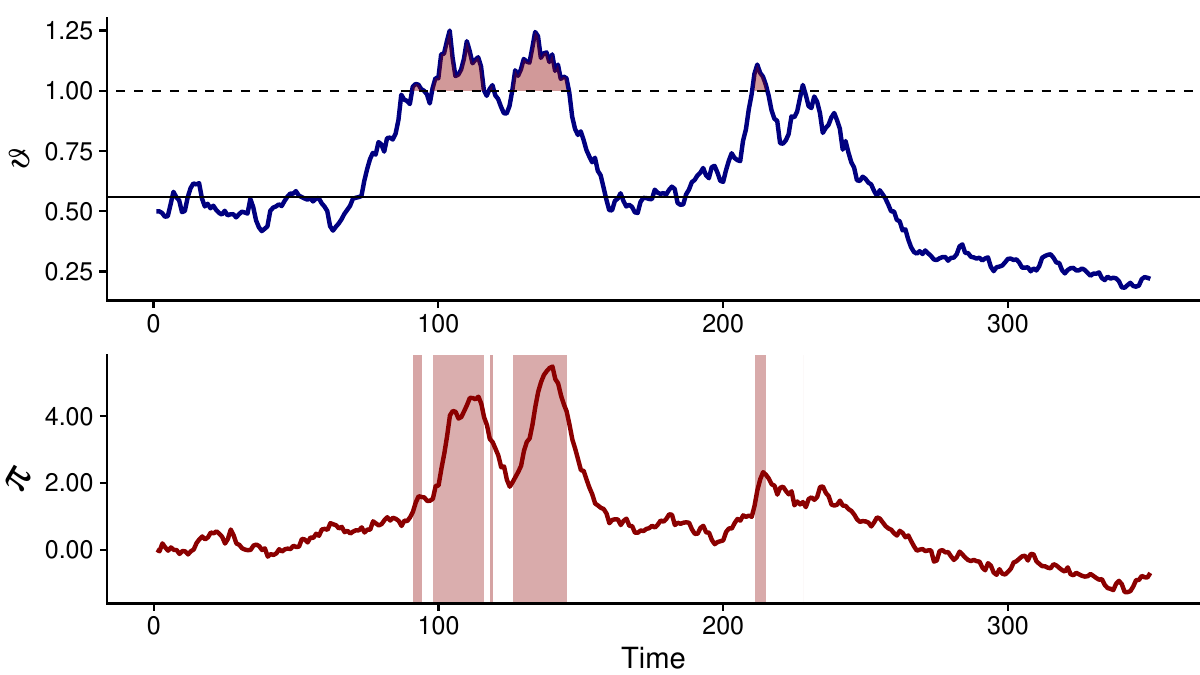}
\begin{minipage}{0.9\textwidth}
\vspace{2pt}
\scriptsize \emph{Note:} We plot a single realization from the \cite{benigno2023s} DGP. The upper panel presents the measure of labor market tightness, $\vartheta_t$, while the lower panel shows the resulting inflation series implied by the nonlinear Phillips curve ($\pi_t$). We indicate $\vartheta_t > 1$ by the red shaded area. The dashed line indicates $\vartheta_t = 1$ and the black solid line shows $\vartheta_t = 0.56$.
\end{minipage}
\end{figure}

This DGP is nonlinear and resembles a threshold model. In particular, as noted, if $\vartheta_t$ exceeds unity, then the slope of the labor market tightness indicator switches. The variance of the shocks to $ \log \vartheta_t$ controls how often this happens in-sample. A realization from the DGP is depicted in Figure \ref{fig:pc_dgp}. In terms of time series dynamics, it matches actual US year-on-year inflation quite well. In around 15 percent of the time, $\vartheta_t$ exceeds one, so nonlinearities kick in.   

For this realization, we estimate BNNs that differ by the number of neurons per layer $Q_\ell = Q \in \{K, \dots, 15\}$, number of layers $L \in \{1, 2, 3\}$, and how we decide on the activation functions. The number of covariates is $K=3$ and consists of the first lag of $\pi_t$, $\log \vartheta_t$ and $\upsilon_t$.  We consider a neuron-specific mixture activation function for the BNN with a single hidden layer and a layer-specific (but not neuron-specific) mixture for cases $L > 1$. Then, we consider a BNN that includes the same activation function across all layers and neurons but we estimate it. And finally, we include a model with a ReLU activation function.  

In what follows, we adopt the following labeling convention for the different models. The activation function is denoted by  $\texttt{flex}$ for a neuron or layer-specific mixture activation function, $\texttt{common}$ for a common mixture activation function across layers and neurons, and $\texttt{ReLU}$ for the ReLU activation function. \texttt{NN-flex} is the BNN that has activation functions that are neuron-specific for the case $L=1$, while for $L>1$ we estimate a single activation function per layer. \texttt{NN-ReLU} is the BNN with ReLU activation function while \texttt{NN-common} uses a common activation function for all layers and neurons.

\begin{figure}[!t]
\caption{Insample fit across layers and neurons. \label{fig:insample_pc}}
\begin{minipage}{0.32\textwidth}
\centering
\scriptsize \texttt{NN-common}
\end{minipage}
\begin{minipage}{0.32\textwidth}
\centering
\scriptsize \texttt{NN-flex}
\end{minipage}
\begin{minipage}{0.32\textwidth}
\centering
\scriptsize \texttt{NN-ReLU}
\end{minipage}

\begin{minipage}{0.32\textwidth}
\centering
\includegraphics[scale=.6]{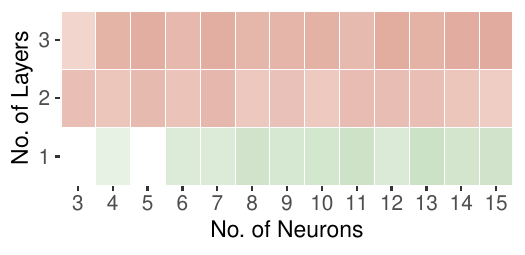}
\end{minipage}
\begin{minipage}{0.32\textwidth}
\centering
\includegraphics[scale=.6]{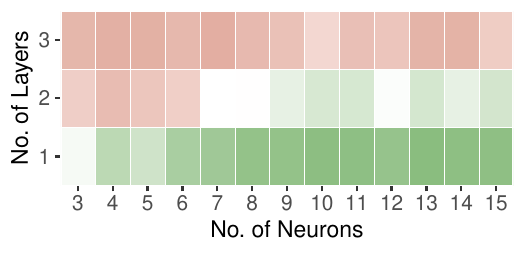}
\end{minipage}
\begin{minipage}{0.32\textwidth}
\centering
\includegraphics[scale=.6]{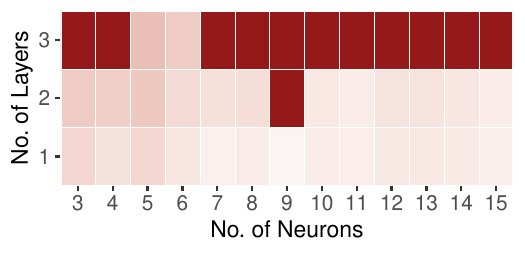}
\end{minipage}

\begin{minipage}{\textwidth}
\centering
\vspace{-10pt}
\includegraphics[scale=.7]{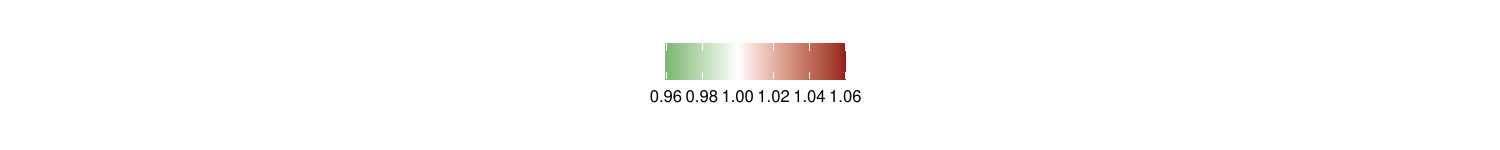}
\end{minipage}

\begin{minipage}{\textwidth}
\vspace{2pt}
\scriptsize \emph{Note:} This figure shows relative root mean squared errors (RMSEs) benchmarked against the simplest BNN considered (i.e., \texttt{NN-common} with a single hidden layer  and three neurons). Values below 1 show outperformance of the benchmark (in green) while inferior specifications give values above 1 (indicated in red). We take averages over 20 random draws from our DGP.
\end{minipage}

\end{figure}

Figure \ref{fig:insample_pc} presents the average across 20 replications of in-sample relative root mean squared errors (RMSEs) from various BNNs benchmarked against the simplest BNN considered (i.e., \texttt{NN-common} with a single hidden layer  and three neurons). Results are summarized in three heatmaps that illustrate the relationship between the number of neurons and layers for different specifications of the activation functions. The left panel shows the results for \texttt{NN-common}. The middle panel includes the results for the BNN that has neuron-/layer-specific activation functions (\texttt{NN-flex}) and the right panel includes \texttt{NN-ReLU}. This allows us to investigate how increasing network complexity (in terms of width, i.e. the number of neurons, and depth, i.e. the number of layers) impacts in-sample estimation accuracy.

From Figure \ref{fig:insample_pc}, we can draw three main conclusions. First, as opposed to other fields, estimation accuracy does not increase when we move from one to two hidden layers. This holds for \texttt{NN-common} and \texttt{NN-flex}, whereas for \texttt{NN-ReLU} the performance remains broadly unchanged. When we move from two to three layers, in all cases estimation accuracy further deteriorates.  This indicates that for a time series such as the one shown in Figure \ref{fig:pc_dgp}, using shallow networks is sufficient and, in fact, recovers implied mean relations with (slightly) more accuracy than if we include more layers.

Second, when we increase the number of neurons we find improvements in estimation  accuracy for all models and most choices of $Q$. This, however, only holds for the first hidden layer when we consider \texttt{NN-common} and \texttt{NN-ReLU}. For \texttt{NN-flex}, it also holds for two hidden layers. Notice that for \texttt{NN-flex}, we find that after including six neurons, performance gains vanish and for $Q \ge 6$, almost all models produce a similar in-sample fit. 

Finally, the single best performing model in this exercise is the \texttt{NN-flex}, with one layer and $Q \in \{10,11,13\}$. Given that in-sample RMSEs do not change appreciably when we include more neurons, using \texttt{NN-flex} with $Q=15$ is a competitive choice that reduces the risk of model mis-specification. Yet, it can increase the risk of overfitting, an issue that we consider in more details in the next section.


\section{Macroeconomic forecasting using BNNs} \label{sec: emp_work}
In this section we show that different BNN models produce accurate density forecasts and we explain why they do so. In the next sub-section we explain the data, forecast design and competing models while Sub-section \ref{sec: emp_results} includes the forecasting results, and Sub-section \ref{sec: properties} zooms into the predictive model properties. 

\subsection{Data overview, competing models, and the forecast exercise design}\label{sec: data}
We use the popular FRED-MD database proposed in \cite{mccracken2016fred}. To gain a comprehensive picture of the importance of nonlinearities in large macro datasets, our forecasting exercise includes the consumer price (CPIAUCSL) inflation rate as specified in \cite{stock1999forecasting}, the month-on-month (m-o-m) growth rate of industrial production (INDPRO), and the m-o-m growth rate of employment (CE16OV).  The sample ranges from January 1960 to December 2020. We assume that these three focus variables are a (unknown) function of the first lag of $K=120$ economic and financial variables. To get an understanding on the relationship between model size and forecast performance, we also consider smaller models that only include as regressor the first lag of the dependent variable and models that include the first eight principal components of the (lagged) variables.\footnote{To determine the number of factors we follow \cite{bai2002determining,bai2013principal} and use the IC$_2$ criterion.} These models can be interpreted as (possibly nonlinear extensions of the) diffusion index regressions in the spirit of \cite{stock2002macroeconomic}.


Our focus is on short-term forecasting. Hence, we compute the one-month-ahead predictive distributions for our hold-out sample, which starts in January 2000 and ends in December 2020 (i.e., $252$ monthly hold-out periods). These forecasts are obtained recursively, meaning that we use the data through January 2000 as a training sample and then forecast one-month-ahead. After obtaining the corresponding predictive densities, the sample is expanded by a single month. This procedure is repeated until the end of the sample is reached. 

We consider different versions of our BNN models. They differ in terms of the number of hidden layers and the choice of the activation function. We consider a deep BNN with $L=3$ hidden layers. Within each layer, we set the number of neurons equal to $K$, and thus include a large number of neurons. Then, we also consider a shallow neural network that includes only a single hidden layer. Within both types of BNNs we consider different versions of our mixture activation function. First, for shallow BNNs we set them to be neuron-specific. For deep BNNs, we estimate the activation function for each layer but not across neurons. Moreover, we consider a deep and a shallow BNN that only uses a single activation function across all layers and neurons that we estimate. The final version is a deep and shallow BNN with ReLU activation function. We state the depth of the network (which can either be \texttt{shallow} or \texttt{deep}) and then the specification for the activation function. For instance, $\texttt{deep-NN-flex}$ is the deep neural network with layer-specific activation functions while $\texttt{shallow-NN-ReLU}$ is the shallow BNN with ReLU activation function. 


We compare the different BNNs to other techniques commonly used in machine learning and applied macroeconometrics. A natural benchmark is the linear regression model. We include several linear models that differ in how regularization is achieved. First, we consider a model that features a horseshoe prior. Then, we include models that use LASSO \citep{tibshirani1996regression} and elastic net regularization \citep{zou2005regularization}, respectively. In addition, we use Bayesian additive regression trees (BART, see e.g. \cite{chipman2010bart} and Sub-section \ref{sec: BART} in the Online Appendix) and random forests \citep[RF,][]{breiman2001random}. Last, we also consider a BNN estimated through back-propagation (labeled BNN-BP). This BNN includes the ReLU activation function and a spike and slab prior on the weights (see Sub-section \ref{sec: BBB} in the Online Appendix for more details). Except for BNN-BP, RF and the elastic net, all models have SV.

To compare models, we rely on log predictive likelihoods (LPLs). Our use of LPLs is motivated by its close relationship to the marginal likelihood, a standard Bayesian measure of model fit \citep[see, e.g.,][]{geweke2010comparing}. Moreover, focusing on point forecast accuracy exclusively implies that we ignore higher-order features of the predictive distributions. 

\subsection{Out-of-sample predictive accuracy}\label{sec: emp_results}
\subsubsection{Overall forecasting performance}
To gain an overall picture of the forecasting performance, we show differences in average LPLs (over the hold-out sample) between a particular model and the linear diffusion index regression model in Table \ref{tab:eval_Macro1_h_1_lpl}. Numbers above zero point towards outperformance of a given model whereas numbers below zero suggest that the benchmark linear model produces more accurate density predictions. 

\begin{table}[!tbp]
{\tiny
\caption{Density forecast performance across 252 hold-out observations.\label{tab:eval_Macro1_h_1_lpl}} 
\begin{center}
\begin{tabular}{lcccccccccccc}
\toprule
\multicolumn{1}{c}{\bfseries Covariates}&\multicolumn{1}{c}{\bfseries }&\multicolumn{11}{c}{\bfseries Model}\tabularnewline
\cline{1-1} \cline{3-13}
\multicolumn{1}{c}{}&\multicolumn{1}{c}{}&\multicolumn{3}{c}{deep-NN-}&\multicolumn{3}{c}{shallow-NN-}&\multicolumn{1}{c}{BNN-BP}&\multicolumn{1}{c}{BART}&\multicolumn{1}{c}{Elastic}&\multicolumn{1}{c}{LASSO}&\multicolumn{1}{c}{Random}\tabularnewline
\multicolumn{1}{c}{}&\multicolumn{1}{c}{}&\multicolumn{1}{c}{common}&\multicolumn{1}{c}{flex}&\multicolumn{1}{c}{ReLU}&\multicolumn{1}{c}{common}&\multicolumn{1}{c}{flex}&\multicolumn{1}{c}{ReLU}&\multicolumn{1}{c}{}&\multicolumn{1}{c}{}&\multicolumn{1}{c}{Net}&\multicolumn{1}{c}{}&\multicolumn{1}{c}{Forest}\tabularnewline
\midrule
\multicolumn{13}{c}{Inflation}\tabularnewline
\multicolumn{13}{l}{\bfseries}\tabularnewline
   AR(1)&   &   \textbf{-0.01}&   -0.02&   -0.01&   -0.02&   -0.01&   -0.02&   -0.19&   -0.22&      -0.04&   -0.04&   -3.63\tabularnewline
      PCA&   &    0.01&   \textbf{0.01}&    0.01&    0.01&    0.01&    0.01&   -0.11&   -0.20&   -0.30&   -0.01&   -0.22\tabularnewline
   Large&   &    0.09&   \textbf{0.09}&    0.09&    0.09&    0.09&    0.08&   -0.12&   -0.03&   -0.21&    0.06&   -0.01\tabularnewline
\midrule
\multicolumn{13}{c}{Industrial Production}\tabularnewline
\multicolumn{13}{l}{\bfseries}\tabularnewline
   AR(1)&   &   \textbf{0.08}&    0.08&    0.08&    0.08&    0.08&    0.08&   -0.81&   -0.12&      0.03&    0.03&   -2.02\tabularnewline
    PCA&   &    0.06&    0.06&    0.06&    0.05&    0.05&   \textbf{0.06}&   -0.46&   -0.01&   -2.55&    0.00&   -0.19\tabularnewline
   Large&   &    0.16&    0.16&   \textbf{0.17}&    0.17&    0.16&    0.16&   -0.41&    0.08&   -1.75&    0.09&   -0.07\tabularnewline
\midrule
\multicolumn{13}{c}{Employment}\tabularnewline
\multicolumn{13}{l}{\bfseries}\tabularnewline
   AR(1)&   &    0.26&    0.24&    0.26&    0.26&   \textbf{0.27}&    0.27&   -0.96&   -0.67&      0.08&    0.08&   -4.83\tabularnewline
      PCA&   &    0.27&   \textbf{0.34}&    0.30&    0.29&    0.32&    0.30&   -0.59&   -0.19&   -6.28&    0.08&   -3.35\tabularnewline
   Large&   &    0.36&   \textbf{0.41}&    0.36&    0.36&    0.38&    0.36&   -0.87&    0.11&   -5.56&    0.18&   -2.49\tabularnewline
\bottomrule
\end{tabular}\end{center}}
\vspace{-10pt}
\centering
\begin{minipage}{0.95\textwidth}
\tiny \emph{Note:} The table shows average log predictive likelihoods (LPLs) relative to the linear benchmark (a diffusion index regression with SV and a Horseshoe prior). In bold we mark the best performing model for each case. Results are averaged across the hold-out. 
\end{minipage}
\end{table}

Starting with inflation forecasts, we observe differences across information sets. In general, using only the first lag of inflation yields density forecasts that are very close to the ones of the linear benchmark model that leverages a larger information set. Within the class of BNNs, there are only small differences across specifications.  When we add the first eight principal components to the lagged inflation (the row labeled PCA in the table) we find small but insignificant gains vis-\'{a}-vis the benchmark for the BNNs. The other machine learning methods, however, are often weaker than the different BNNs, with LASSO the best performing among them. This is not surprising given that the PCA regression is estimated under a horseshoe prior, and both the horseshoe and LASSO are related, in the sense that the LASSO can be interpreted as a Bayesian global-local shrinkage prior. 

If we include all variables in an unrestricted manner, results change somewhat. The different BNNs  outperform the benchmark.  We find only small differences among the BNNs, with deep and shallow models producing very similar density forecasts. When we compare different specifications for the activation functions, we find no discernible differences (except for a slightly weaker performance for the shallow BNN with ReLU activation functions). In terms of the competing models, we find LASSO-based density predictions to remain competitive, whereas those from the other models are inferior to the benchmark.   

Next, we consider forecasts for industrial production (IP) growth. Starting again with the AR(1) information set, we find that all BNN-based models produce LPLs that are approximately the same, and outperform the benchmark. The fact that predictive accuracy is very similar across BNNs is not surprising, since we only leverage information embodied in lagged IP growth.  Most other machine learning methods do not improve upon the benchmark model, except for the LASSO (and elastic net, which however does worse with a larger information set). When we add more information, we find a similar pattern, with slightly larger gains for the BNNs, rather similar across specifications.  


Finally, we consider employment growth forecasts. The results mirror those for inflation and IP growth. Neural networks, across information sets, perform well and improve upon all competing machine learning specifications by appreciable margins. Interestingly, we find more variation with respect to how we treat the activation function. In principle, if we use our mixture specification and make the activation function either layer or neuron-specific (in the case of the shallow learners), we almost always gain relative to the model that fixes the activation functions to be of ReLU type. 


To sum up, different variants of the BNNs we propose outperform most other common machine learning models in terms of quality of short-term density forecasts for key US economic indicators. Differences between deep and shallow BNNs are rather small, indicating that for practitioners interested in predicting output, inflation or employment, a shallow BNN that is easier to handle is a good choice that delivers accurate density forecasts on average. While our mixture activation function only translates into modest gains in terms of density forecasting performance, it is worth stressing that it frees the researcher from the necessity to decide on one particular activation function and thus reduces the number of inputs to the model. We should also stress that in most cases, using more information in an unrestricted manner pays off. This finding, however, is not specific to all models we consider but mostly to BNNs.


\subsubsection{Forecasting performance over time}
\begin{figure}[t!]
\caption{Cumulative LPLs for BNNs and the best performing benchmark. \label{fig:oos_lps}}
\begin{minipage}{0.32\textwidth}
\centering
\scriptsize \textit{Inflation}
\end{minipage}
\begin{minipage}{0.32\textwidth}
\centering
\scriptsize \textit{Industrial production}
\end{minipage}
\begin{minipage}{0.32\textwidth}
\centering
\scriptsize \textit{Employment}
\end{minipage}

\begin{minipage}{0.32\textwidth}
\centering
\includegraphics[scale=.4]{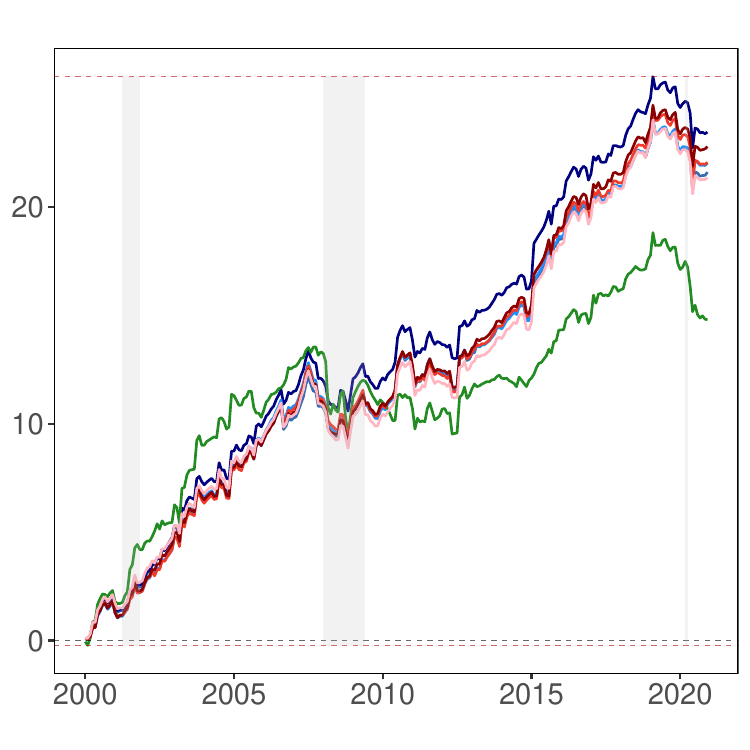}
\end{minipage}
\begin{minipage}{0.32\textwidth}
\centering
\includegraphics[scale=.4]{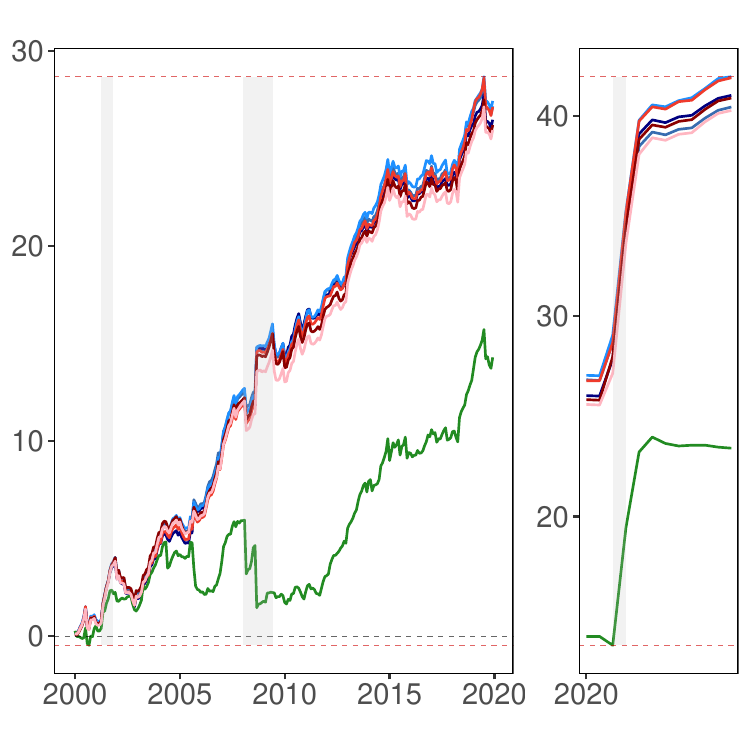}
\end{minipage}
\begin{minipage}{0.32\textwidth}
\centering
\includegraphics[scale=.4]{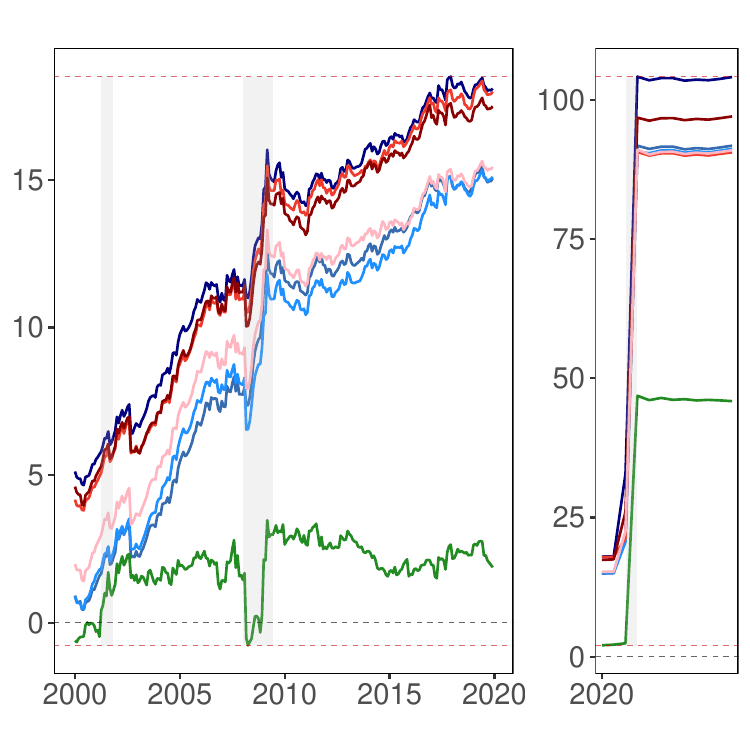}
\end{minipage}

\begin{minipage}{\textwidth}
\centering
\vspace{-0.8cm}
\includegraphics[scale=.4]{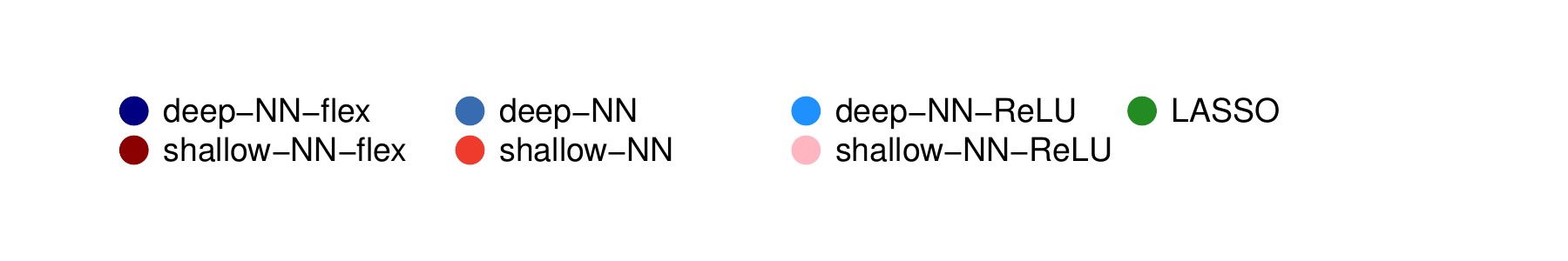}
\end{minipage}

\begin{minipage}{\textwidth}
\vspace{2pt}
\scriptsize \emph{Note:} We plot the evolution of cumulative log predictive likelihoods (LPLs) against the linear diffusion index model for the one-step-ahead predictive densities.   The red dashed lines denote the max./min. LPLs, while the gray shaded areas indicate the NBER recessions. 
\end{minipage}
\end{figure}

The discussion in the previous section focused on how well the different models perform in terms of overall forecast accuracy for the full hold-out period. To drill into performance differences over time,  Figure \ref{fig:oos_lps} shows cumulative one-step-ahead LPLs relative to the linear benchmark model over time, with positive values indicating that the nonlinear models are better than the linear benchmark.  Given their strong performance, we focus on the models that feature the large dataset and only include the LASSO as the best performing approach among the competitors. 

At a general level, we find that modeling nonlinearities pays off during turbulent times such as the recession in the early 2000s, the GFC and during the pandemic. Only early in the GFC, the linear benchmark produces slightly more accurate density forecasts.\footnote{Notice that this does not hold for the LASSO. We conjecture that this is driven by the fact that the LASSO tends to overshrink significant signals.} But halfway through the GFC this pattern shifts and most nonlinear learners outperform linear models. 

For IP and employment growth we find substantial gains of nonlinear models during recessionary episodes. Notice that for these two focus variables, we have added separate panels that show the performance throughout the pandemic. Focusing on these we find heavy gains in early 2020. These gains persist but flatten out afterwards.   This observation provides evidence that a high degree of model flexibility, and thus capturing these severe outliers in real economic activity, pays off in terms of predictive accuracy. Such large gains in forecast accuracy are not evident for inflation, which remained rather stable relative to industrial production and employment growth in 2020 and did not exhibit these severe outliers. 

Consistent with the overall findings, deep and shallow models display a similar performance, making it difficult to find systematic differences over time. 
Yet, ReLU-based models are consistently weaker than their counterparts with a mixture activation function for employment. It is also interesting that deep BNNs with ReLU activation functions are beaten by shallow BNNs with flexible activation functions. This pattern holds for all three focus variables and seems to be consistent over time.\footnote{Figure \ref{fig:fluct_test} in the Online Appendix reports the fluctuation test statistic relative to the benchmark as proposed by \cite{giacomini2010forecast}. It turns out that the gains with respect to the benchmark are present and statistically significant for all the BNN models, particularly so for inflation and industrial production, while they are only occasionally significant for LASSO.}

\subsubsection{The role of stochastic volatility}

\begin{table}[h!]
{\tiny
\caption{Density forecasting performance of the various NNs without SV in the shocks.\label{tab:eval_FCST-Macro1_SVFALSE}} 
\begin{center}
\begin{tabular}{lccccccc}
\toprule
\multicolumn{1}{c}{\bfseries Covariates}&\multicolumn{1}{c}{\bfseries }&\multicolumn{6}{c}{\bfseries Model}\tabularnewline
\cline{1-1} \cline{3-8}
\multicolumn{1}{c}{}&\multicolumn{1}{c}{}&\multicolumn{3}{c}{deep-NN-}&\multicolumn{3}{c}{shallow-NN-}\tabularnewline
\multicolumn{1}{c}{}&\multicolumn{1}{c}{}&\multicolumn{1}{c}{common}&\multicolumn{1}{c}{flex}&\multicolumn{1}{c}{ReLU}&\multicolumn{1}{c}{common}&\multicolumn{1}{c}{flex}&\multicolumn{1}{c}{ReLU}\tabularnewline
\midrule
\multicolumn{8}{c}{Inflation}\tabularnewline
\multicolumn{8}{l}{\bfseries}\tabularnewline
   AR(1)&   &   -0.26&   \textbf{-0.26}&   -0.26&   -0.26&   -0.26&   -0.26\tabularnewline
   PCA&   &   -0.21&   \textbf{-0.13}&   -0.21&   -0.21&   -0.21&   -0.20\tabularnewline
   Large&   &   -0.07&   \textbf{-0.04}&   -0.06&   -0.07&   -0.06&   -0.07\tabularnewline
\midrule
\multicolumn{8}{c}{Industrial production}\tabularnewline
\multicolumn{8}{l}{\bfseries}\tabularnewline
   AR(1)&   &   -0.88&   -0.88&   \textbf{-0.84}&   -0.89&   -0.99&   -0.89\tabularnewline
   PCA&   &   -1.66&   \textbf{-0.07}&   -2.06&   -1.92&   -2.35&   -1.38\tabularnewline
   Large&   &   -0.73&   \textbf{0.03}&   -0.37&   -0.54&   -0.58&   -0.63\tabularnewline
\midrule
\multicolumn{8}{c}{Employment}\tabularnewline
\multicolumn{8}{l}{\bfseries}\tabularnewline
   AR(1)&   &    -4.97&    -4.72&   \textbf{-4.56}&    -4.60&   -16.11&    -5.26\tabularnewline
   PCA&   &   -6.56&   \textbf{-2.42}&   -5.10&   -6.82&   -9.87&   -9.28\tabularnewline
   Large&   &   -5.93&   \textbf{0.28}&   -5.88&   -5.29&   -5.83&   -6.05\tabularnewline
\bottomrule
\end{tabular}\end{center}}
\vspace{-10pt}
\centering
\begin{minipage}{0.58\textwidth}
\tiny \emph{Note:} The table shows average log predictive likelihoods (LPLs) relative to the linear benchmark (a diffusion index regression with SV and a Horseshoe prior). In bold we mark the best performing model for each case. Results are averaged across the hold-out. 
\end{minipage}
\end{table}

One of our additions to the standard NN toolkit is the introduction of SV. To investigate the empirical relevance of this, we now re-do the forecast exercise but turn off SV for the BNNs. The results of this exercise are shown in Table \ref{tab:eval_FCST-Macro1_SVFALSE}. The table again shows the LPLs relative to the linear diffusion index model with SV. Hence, LPL differences between Table \ref{tab:eval_Macro1_h_1_lpl} and Table \ref{tab:eval_FCST-Macro1_SVFALSE} are directly comparable.

Turning off SV hurts density forecasting performance. In most cases, LPL differences are lower than the ones observed in Table \ref{tab:eval_Macro1_h_1_lpl} by varying margins.  In fact, for inflation, we find that the benchmark linear model with SV improves upon \texttt{deep-NN-flex} and \texttt{shallow-NN-flex} with the large dataset.  There are only small differences in terms of forecast accuracy between the deep and shallow models. This is consistent with the findings based on the models that use SV.

For IP and employment growth, we find that \texttt{deep-NN-flex} is capable of  improving upon the benchmark. These gains are muted for IP and slightly more pronounced for employment.  What is interesting, however, is that the choice of the activation function seems to matter much more when we use homoskedastic BNNs. The layer-specific mixture activation function in the case of a deep BNN yields much more precise density forecasts than a ReLU-based BNN and a specification with a common mixture activation function. Notice, however, that for IP growth we observe that deep BNNs seem to cope much better with model mis-specification in terms of the shock volatility processes than shallow models.

\subsection{Predictive model properties}\label{sec: properties}
To better understand the predictive performance of our different BNNs, we now consider two specifications and zoom into specific model features. The first is the deep BNN with layer-specific activation functions (\texttt{deep-NN-flex}) and the second one is the shallow BNN with neuron-specific activation functions (\texttt{shallow-NN-flex}). 
\subsubsection{Which activation function?}
A natural starting point is the specific form of the activation function if we use our mixture specification.  To investigate which activation function receives more model weight, we focus on how much weight each activation function attains under the posterior. Recall that the mixture activation function mixes over four different activation functions (leakyReLU, sigmoid, ReLU and tanh). Table \ref{tab:actfunc_FCST-Macro1} shows the average of the posterior estimates of $\omega_{\ell, q}^{(m)}$ associated with each of the four activation functions over the hold-out period.  The rows show the different layers and, to simplify the table, we have averaged over the weights associated with the different neurons in the case of \texttt{shallow-NN-flex}.  

\begin{table}[!tbp]
{\tiny
\caption{Posterior weights of activation functions.\label{tab:actfunc_FCST-Macro1}} 
\begin{center}
\begin{tabular}{lccccccccccccc}
\toprule
\multicolumn{1}{c}{\bfseries Layer}&\multicolumn{1}{c}{\bfseries }&\multicolumn{4}{c}{\bfseries Inflation}&\multicolumn{4}{c}{\bfseries Industrial production}&\multicolumn{4}{c}{\bfseries Employment}\tabularnewline
\cline{1-1} \cline{3-14}
\multicolumn{1}{c}{}&\multicolumn{1}{c}{}&\multicolumn{1}{c}{leakyReLU}&\multicolumn{1}{c}{ReLU}&\multicolumn{1}{c}{sigmoid}&\multicolumn{1}{c}{tanh}&\multicolumn{1}{c}{leakyReLU}&\multicolumn{1}{c}{ReLU}&\multicolumn{1}{c}{sigmoid}&\multicolumn{1}{c}{tanh}&\multicolumn{1}{c}{leakyReLU}&\multicolumn{1}{c}{ReLU}&\multicolumn{1}{c}{sigmoid}&\multicolumn{1}{c}{tanh}\tabularnewline
\midrule
\multicolumn{14}{c}{\texttt{shallow-NN-flex}}\tabularnewline
\multicolumn{14}{l}{\bfseries}\tabularnewline
    1&   &   23.9&   23.8&   26.2&   26.1&   23.8&   23.8&   26.3&   26.1&   23.7&   23.7&   26.0&   26.6\tabularnewline
\midrule
\multicolumn{14}{c}{\texttt{deep-NN-flex}}\tabularnewline
\multicolumn{14}{l}{\bfseries}\tabularnewline
    1&   &   24.6&   24.0&   23.4&   28.0&   24.9&   23.8&   23.1&   28.2&   21.2&   19.9&   19.4&   39.5\tabularnewline
    2&   &   24.2&   24.1&   23.6&   28.1&   24.1&   23.9&   24.9&   27.1&   25.8&   21.2&   21.0&   32.0\tabularnewline
    3&   &   ~3.5&   ~3.4&   44.3&   48.8&   ~3.6&   ~3.6&   45.2&   47.6&   ~3.6&   ~3.6&   38.3&   54.5\tabularnewline
\bottomrule
\end{tabular}\end{center}}
\vspace{-10pt}
\centering
\begin{minipage}{\textwidth}
\tiny \emph{Note:} The table shows mean posterior estimates of $\omega_{\ell, q}^{(m)}$ for each layer averaged across the hold-out periods. For \texttt{shallow-NN-flex}, we average over the weights of all neurons. All numbers are in percentages.
\end{minipage}
\end{table}

For the shallow model, we find that all activation functions receive almost equal weights across the three focus variables. Only sigmoid and tanh obtain slightly more posterior weight. This feature is more pronounced for employment, but differences are quite small. 

For \texttt{deep-NN-flex} more interesting patterns arise. In this case, we find more asymmetries in terms of activation functions across variables and layers. For inflation, tanh receives slightly more posterior weight for the first two hidden layers. In the output layer, this pattern shifts and leakyReLU and ReLU attain only little weight while sigmoid and tanh obtain over 90 percent of total posterior weight. For IP, a similar pattern shows up. In the first two layers, we find posterior weights that are more symmetric while in the output layer, sigmoid and tanh dominate and leakyReLU and ReLU only play a limited role.

For employment growth, the variable for which we find the strongest gains, the pattern is different. For layers one and two, tanh receives the largest posterior weight with the remaining three activation functions having similar mean estimates of $\omega_{\ell, q}^{(m)}$. For the output layer, we again find that sigmoid becomes substantially more important, receiving a weight of around 38 percent while tanh still dominates and receives over 54 percent of posterior weights.

This brief discussion shows that in the predictive exercise, our BNNs do not generate strong forecasts by relying on a single activation function but by averaging over all four of them (in particular for the deep model and hidden layers below the output layer). In the case of the deep BNN, the model places substantial posterior weight on sigmoid and tanh, a pattern that also shows up for the shallow model but in a much more attenuated manner.


\subsubsection{The relationship between in-sample fit and out-of-sample predictability}\label{sec: insample}
The results up to this point tell a story that flexible models help in extreme periods and are competitive in normal times.  We now ask whether neural networks extract information from $\bm x_t$ that linear models can not exploit and how this impacts predictive accuracy (measured again through LPLs). 

\begin{figure}[htb!]
\caption{Relative R2 against relative LPL. \label{fig:scatter_lpl_r2}}
\centering
\begin{minipage}{0.49\textwidth}
\centering
\scriptsize \texttt{shallow-NN-flex}
\end{minipage}
\begin{minipage}{0.49\textwidth}
\centering
\scriptsize \texttt{deep-NN-flex}
\end{minipage}
\begin{minipage}{1\textwidth}
    \centering (a) \textit{Inflation}
\end{minipage}
\begin{minipage}{0.49\textwidth}
\centering
\includegraphics[scale=.3]{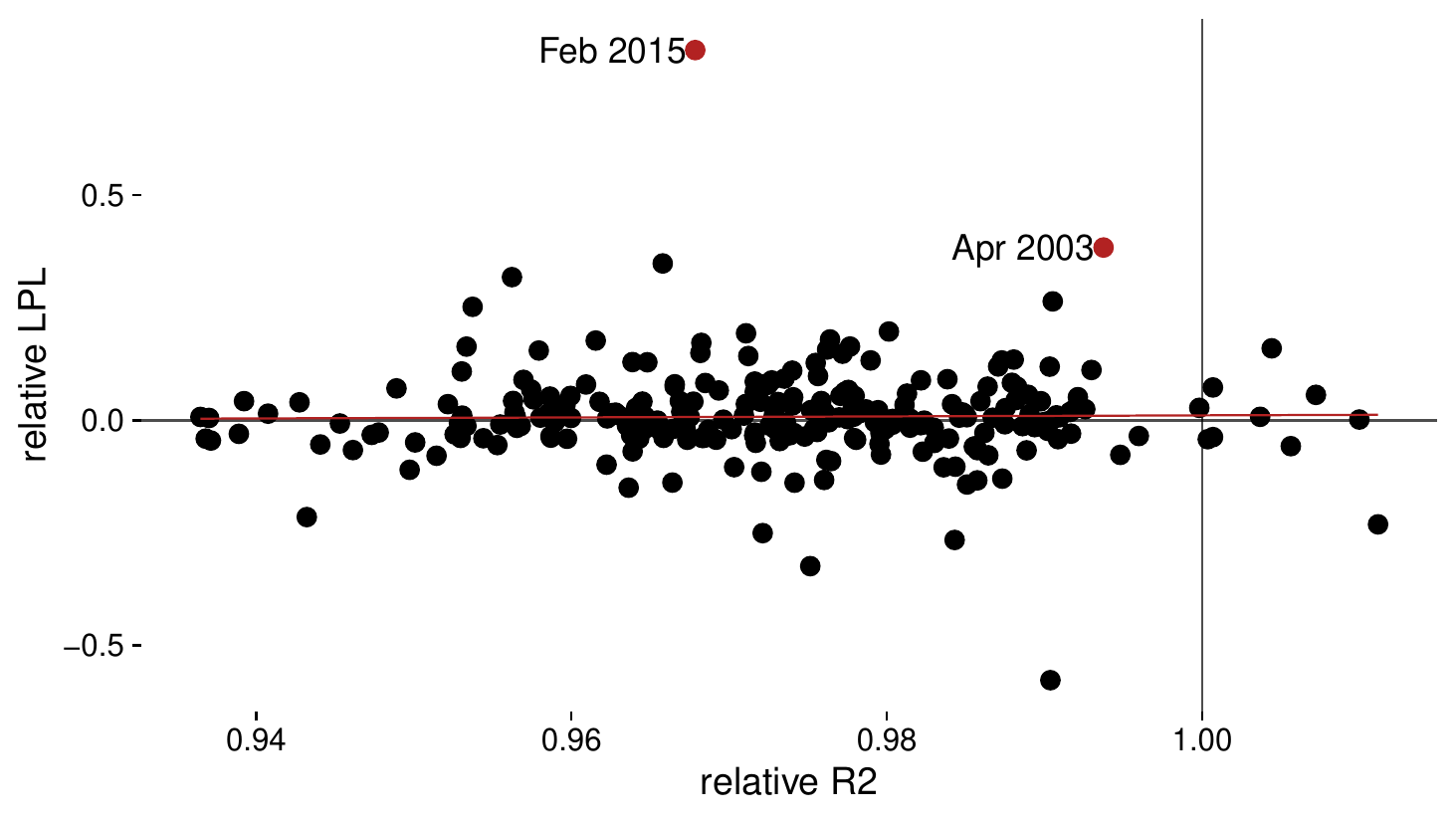}
\end{minipage}
\begin{minipage}{0.49\textwidth}
\centering
\includegraphics[scale=.3]{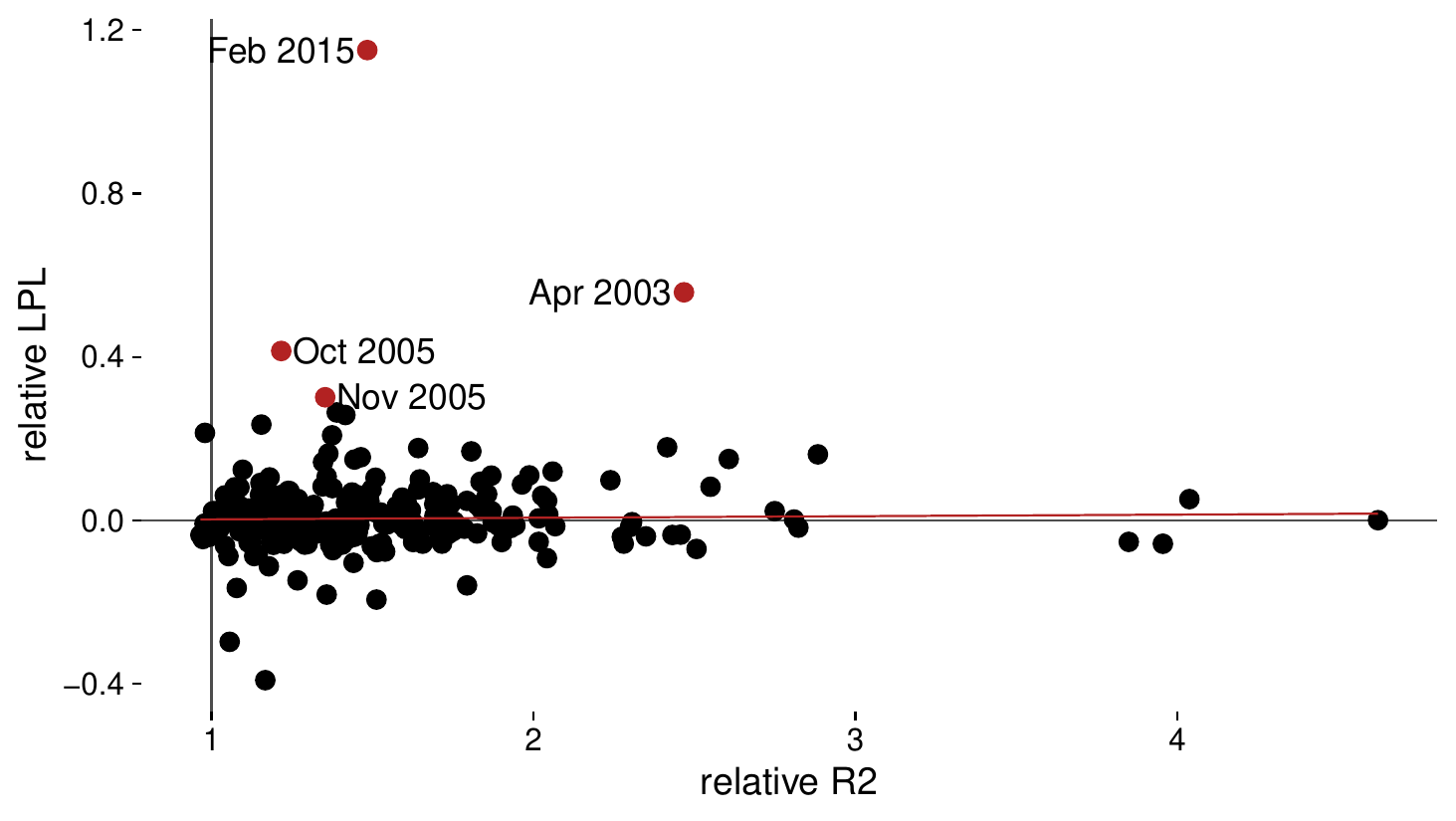}
\end{minipage}

\vspace{10pt}
\centering
\begin{minipage}{1\textwidth}
\centering
(b) \textit{Industrial production}
\end{minipage}

\begin{minipage}{0.49\textwidth}
\centering
\includegraphics[scale=.3]{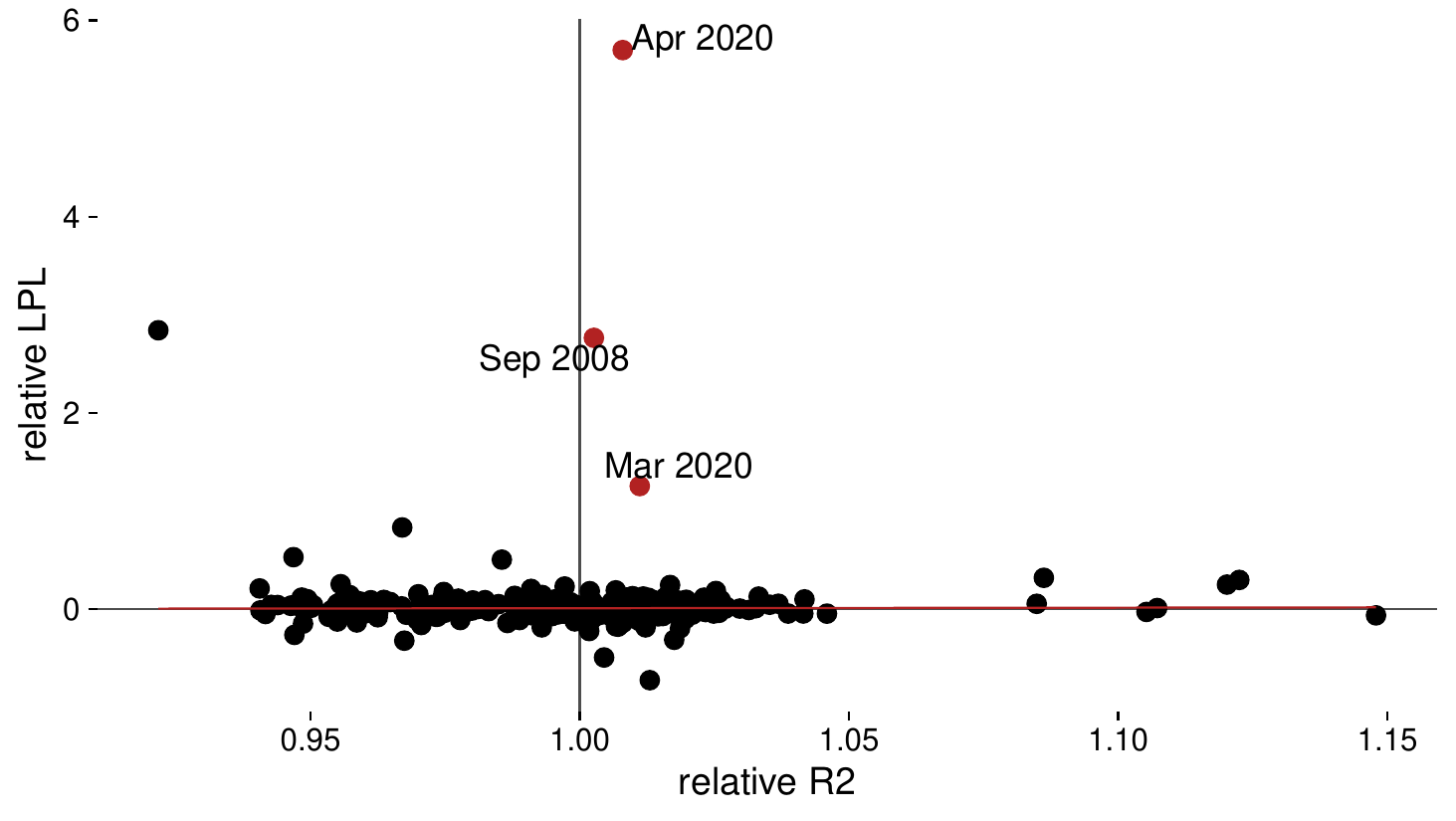}
\end{minipage}
\begin{minipage}{0.49\textwidth}
\centering
\includegraphics[scale=.3]{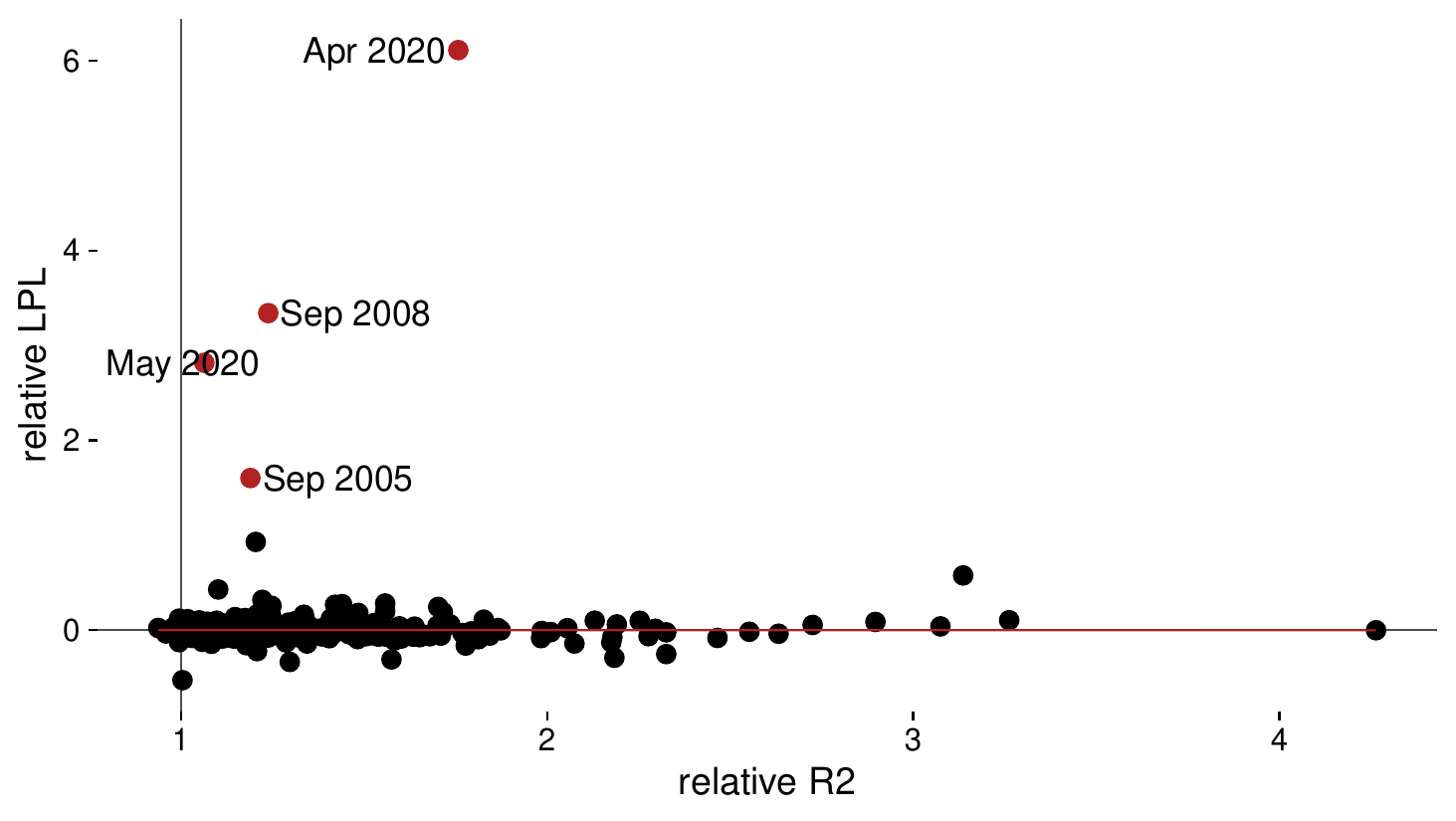}
\end{minipage}
\vspace{10pt}
\centering
\begin{minipage}{1\textwidth}
\centering
(c) \textit{Employment}
\end{minipage}

\begin{minipage}{0.49\textwidth}
\centering
\includegraphics[scale=.3]{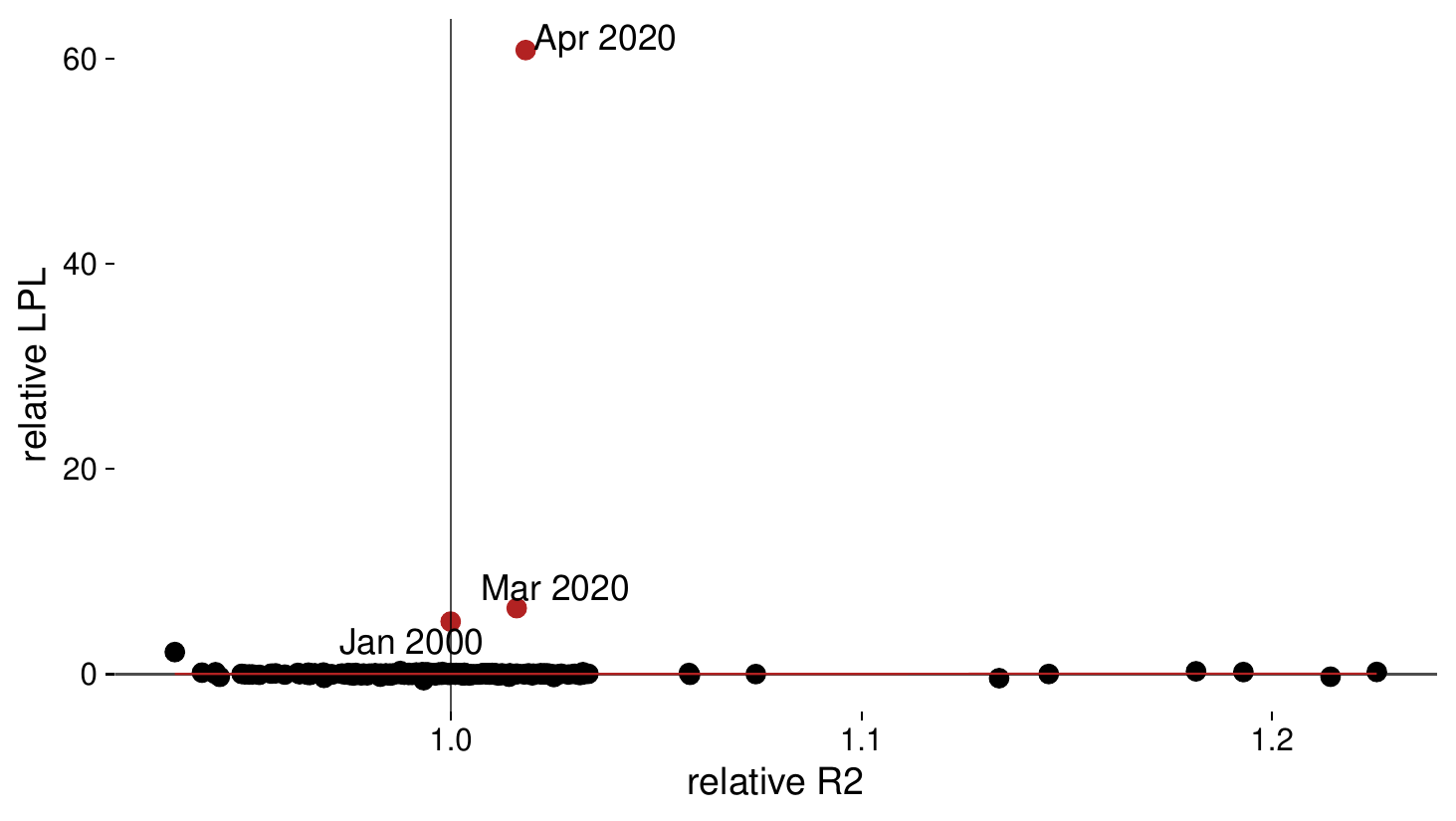}
\end{minipage}
\begin{minipage}{0.49\textwidth}
\centering
\includegraphics[scale=.3]{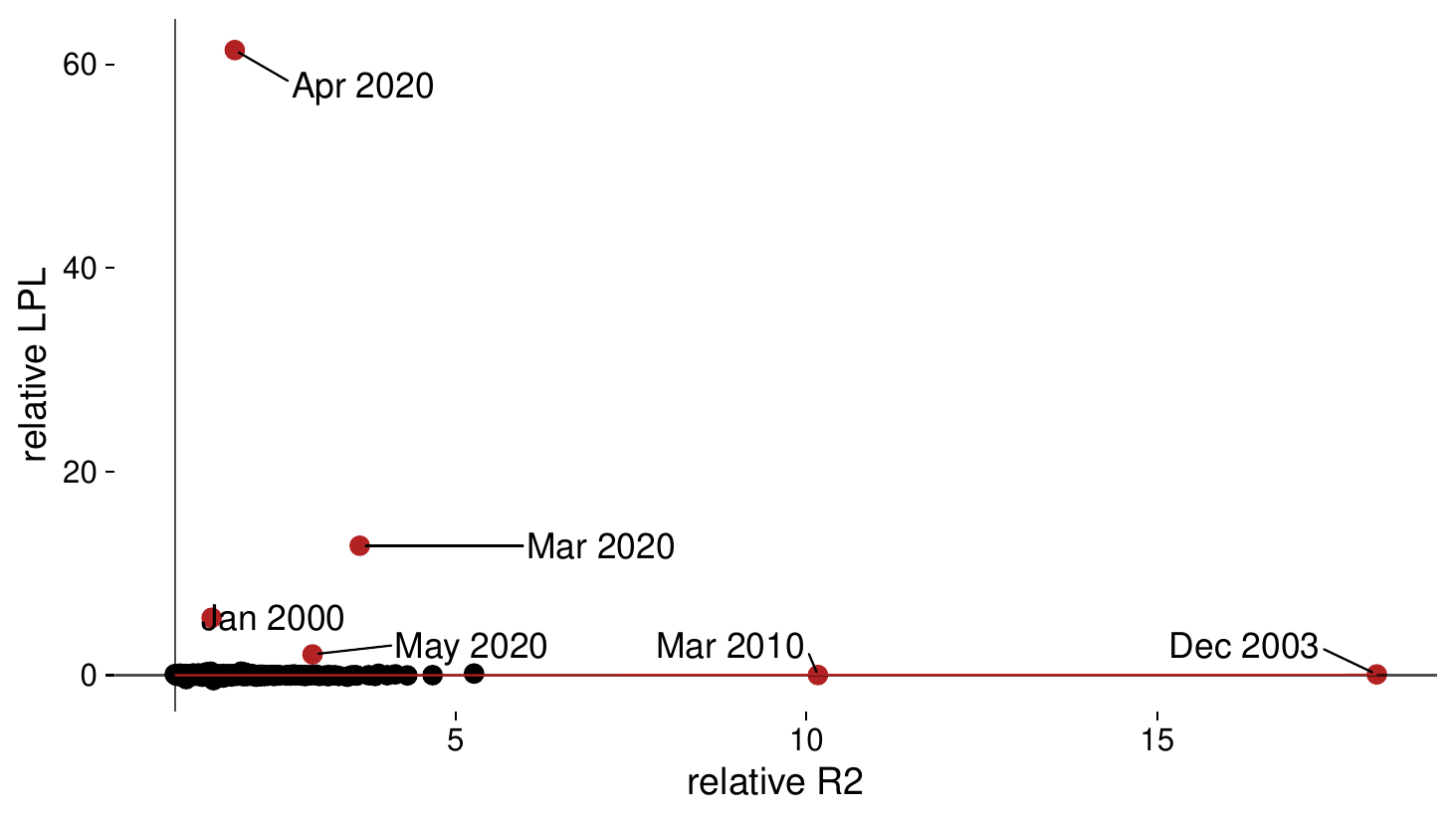}
\end{minipage}

\begin{minipage}{\textwidth}
\vspace{2pt}
\scriptsize \emph{Note:} This figure shows relative R2 versus relative LPL of \texttt{shallow-NN-flex} and \texttt{deep-NN-flex} against the linear model for each of our three applications. We color observations which feature high in-sample fit and high predictive accuracy relative to the benchmark.
\end{minipage}
\end{figure}

To achieve this, we compute the amount of variation explained through the conditional mean piece (labeled R2) for \texttt{shallow-NN-flex} and \texttt{deep-NN-flex}. These R2s are computed recursively and put in relation to the corresponding $t$-by-$t$ LPL using a simple scatter plot. The scatter plots are provided in Figure \ref{fig:scatter_lpl_r2}. The horizontal line at zero implies that if a point is below zero, the linear regression model produces superior density forecasts whereas in the opposite case, the BNN is forecasting better. Points to the left of the vertical line (which stands at one) imply less explanatory power of the BNN whereas points to the right indicate that the conditional mean part of the BNN explains more of the variation in the response as the linear model.   

Two examples illustrate how the scatter plot can be interpreted. Points in the quadrant with $\text{R2}>1$ and $\text{LPL}>0$ represent situations where the BNN is extracting information from $\bm x_t$  that leads to a higher in-sample fit and this information translates into more accurate density forecasts. If $\text{R2}\approx1$ but $\text{LPL}>0$ both models explain a similar amount of in-sample variation but density forecasting performance of the BNN is superior. In this case, these differences are likely driven by higher order moments of the predictive distribution.

Starting with inflation (see panel (a) of the figure) reveals two interesting differences across shallow and deep learners. The shallow neural network produces competitive forecasts based on an in-sample fit that is comparable to the linear diffusion index regression whereas \texttt{deep-NN-flex} produces density forecasts of similar quality but does so with a much stronger in-sample fit that is almost always higher than the one of the linear model, with relative R2 frequently exceeding 1.5.  

For IP growth (depicted in panel (b)), we observe a similar pattern. Shallow models produce in-sample fits which are close to the ones of the linear model. By contrast, the R2-LPL combinations for the deep BNNs are mostly located in the top right quadrant. In both cases, however, BNNs produce IP forecasts in the pandemic (April 2020) which are much more precise than the ones of the linear model and do so while explaining more of the in-sample fit. For \texttt{deep-NN-flex} we also find that density predictions for output growth are much more precise during September 2008, the month of the collapse of Lehman Brothers. 

The overall story that shallow BNNs produce accurate density forecasts with an in-sample fit that is close to the one of the linear model while deep models do the same but with a much higher in-sample fit carries over to employment growth predictions in panel (c). For shallow BNNs we find that around half of the points are above one in terms of relative R2 and for deep BNNs this share is much higher. In both cases, nonlinear models heavily outperform the linear benchmark in April 2020 and, to a lesser degree, in March 2020. In both periods, employment numbers exhibited historic declines due to the Covid-19 pandemic.

\begin{table}[!tbp]
{\tiny
\caption{Measures for the in-sample and out-of-sample outperformance of the BNN against the linear benchmark.\label{tab:r2_lpl}} 
\begin{center}
\begin{tabular}{lcccccc}
\toprule
\multicolumn{1}{l}{\bfseries }&\multicolumn{2}{c}{\bfseries Inflation}&\multicolumn{2}{c}{\bfseries Industrial Production}&\multicolumn{2}{c}{\bfseries Employment}\tabularnewline
\cline{2-7}
\multicolumn{1}{l}{\bfseries }&\multicolumn{1}{c}{shallow-NN-}&\multicolumn{1}{c}{deep-NN-}&\multicolumn{1}{c}{shallow-NN-}&\multicolumn{1}{c}{deep-NN-}&\multicolumn{1}{c}{shallow-NN-}&\multicolumn{1}{c}{deep-NN-}\tabularnewline
\multicolumn{1}{l}{\bfseries }&\multicolumn{1}{c}{flex}&\multicolumn{1}{c}{flex}&\multicolumn{1}{c}{flex}&\multicolumn{1}{c}{flex}&\multicolumn{1}{c}{flex}&\multicolumn{1}{c}{flex}\tabularnewline
\midrule
  \bfseries Outperformance  & 54.0 & 48.8 & 54.0  & 46.4 & 55.6 & 53.6  \tabularnewline
 \bfseries Outperformance \& higher R2 & 2.0  & 48.4 & 26.6  & 45.2&  25.0  & 53.2 \tabularnewline
\bottomrule
\end{tabular}\end{center}}
\vspace{-10pt}
\centering
\begin{minipage}{0.88\textwidth}
\tiny \emph{Note:} The table shows two different measures describing the relationship between in-sample fit and out-of-sample predictability. The measure ``Outperformance" gives the share of relative LPL being positive (i.e., corresponding to the datapoints in the upper quadrants of Figure \ref{fig:scatter_lpl_r2}). The second measure ``Outperformance \& higher R2" gives the share of relative LPL being positive and a relative R2 above one (i.e., corresponding to the datapoints in the right-upper quadrants of Figure \ref{fig:scatter_lpl_r2}). All numbers are in percentages.
\end{minipage}
\end{table}

Table  \ref{tab:r2_lpl} summarizes these findings in a few numbers. The first row, labeled 'Outperformance' shows the number of times a particular BNN improves upon the linear benchmark, the second row, labeled 'Outperformance \& higher R2' gives the percentage of times a model improves upon the linear model and does so with a higher in-sample R2.   This table shows that shallow models have a slightly higher share of outperformance than deep models. For \texttt{shallow-NN-flex}, this outperformance is often achieved with a lower R2 than the benchmark (two percent for inflation; 27 percent for IP and 25 percent for employment). By contrast, deep BNNs frequently outperform the benchmark with higher R2 measures. This pattern points towards benign overfitting, commonly found in the literature \citep[see, e.g., ][]{bartlett2020benign}, but it also suggests that, as far as density forecasting inflation, industrial production and employment is considered, shallow BNNs can be preferred to deep BNNs, as differences in density forecasts are minor and the computational costs are much smaller.

\section{Structural inference with BNNs: The nonlinear effects of financial shocks}\label{sec: structural analysis}
We have shown that BNNs can be used to produce more accurate density forecasts than linear models and other machine learning methods.  This is important for policy making since it immediately implies that BNNs can be used for projection exercises or used as benchmarks for structural models. However, policy makers often have a keen interest in how economic shocks dynamically impact the economy. Our BNNs are well suited for this task and we illustrate it by means of a topical example. We focus on how financial shocks impact inflation, output and employment, and whether these effects are symmetric and proportional. In the literature, there is substantial evidence of asymmetries between benign and adverse financial shocks \citep[see, e.g.,][]{balke2000credit,brunnermeier2014macroeconomic,barnichon2022effects}.

To shed light on this issue, we develop nonlinear local projections \citep[LPs; see, e.g.,][]{mumtaz2022impulse, gonccalves2024state, inoue2024local} for  BNNs. Let $\zeta_t$ denote an exogenous instrument for a shock of interest. Adding $\zeta_t$ to our general nonlinear regression problem and shifting $y_t$ $h$-periods forward (for $h=0,\dots, H$), yields:
\begin{equation*}
    y_{t+h} =  \psi_h \zeta_t + \bm x_t' \bm \gamma_h + f_h(\zeta_t, \bm x_t) + \epsilon_{t+h}.
\end{equation*}
Note that $\psi_h, \bm \gamma_h$ and $f_h$ are horizon-specific. The $h$-step-ahead effect of a unit increase in $\zeta_t$ on $y_{t+h}$ is given by:
\begin{equation*}
    \frac{\partial y_{t+h}}{ \partial \zeta_t} = \psi_h + \frac{\partial f_h(\zeta_t, \bm x_t)}{\partial \zeta_t}.
\end{equation*}
The term  ${\partial f_h(\zeta_t, \bm x_t)}/{\partial \zeta_t}$ depends on $\bm x_t$, indicating that the $h-$step-ahead local projection is dependent on the current state of a set of control variables $\bm x_t$. We integrate out the effects of $\bm x_t$ by approximating the nonlinear local projections with:
\begin{equation*}
    \frac{\partial y_{t+h}}{ \partial \zeta_t} \approx \mathbb{E}(y_{t+h}|\zeta_t=\tau, \bm x_t) - \mathbb{E}(y_{t+h}|\zeta_t = 0, \bm x_t).
\end{equation*}
This represents the difference between the h-step-ahead predictive density conditional on shock $\tau$ of a particular size and sign, and the unconditional h-step-ahead predictive density. This quantity can be computed for each MCMC draw  of $\psi_h, \bm \gamma_h$ and $\widehat{f}_h$ (the trained BNN approximation to $f_h$).

As noted in, e.g., \cite{jorda2005estimation}, LP regressions ignore serial correlation in the forecast errors. In an extensive robustness check, \cite{clark2024investigating} show that the neglect of serial correlation in the shocks only has a small impact on direct forecasts. However, here we follow \cite{lusompa2023local} and include estimates of $\varepsilon_t, \dots, \varepsilon_{t+h-1}$ in $\bm x_t$ to induce serial correlation in the forecast errors.  

In this section, we use the \texttt{shallow-NN-flex} model but simplify it by including only a proxy of a financial shock, and the first lag of $y_t$ as covariates. The proxy of the financial shock $\zeta_t$ is measured through the excessive bond premium \citep[EBP,][]{gilchrist2012credit}. To obtain a shock measure related to EBP, we first consider estimating a structural vector autoregressive (VAR) model, which is consistent with \cite{gilchrist2012credit}. In addition to our three target variables, this VAR model also includes the federal funds rate and stock market returns. We order the EBP measure first and obtain the structural economic shock related to this variable, which can be interpreted as a financial shock then.

\begin{figure}[th!]
\caption{Nonlinear local projections to financial shocks. \label{fig:shlwBNN_lp}}
\centering
\begin{minipage}{0.49\textwidth}
\centering
\normalsize (a) \textit{Sign asymmetries}
\end{minipage}
\begin{minipage}{0.49\textwidth}
\centering
\normalsize (b) \textit{Size asymmetries}
\end{minipage}
\begin{minipage}{1\textwidth}
\centering
\vspace*{10pt}
\small \textit{Inflation}
\vspace*{10pt}
\end{minipage}
\begin{minipage}{0.49\textwidth}
\centering
\includegraphics[scale=.55]{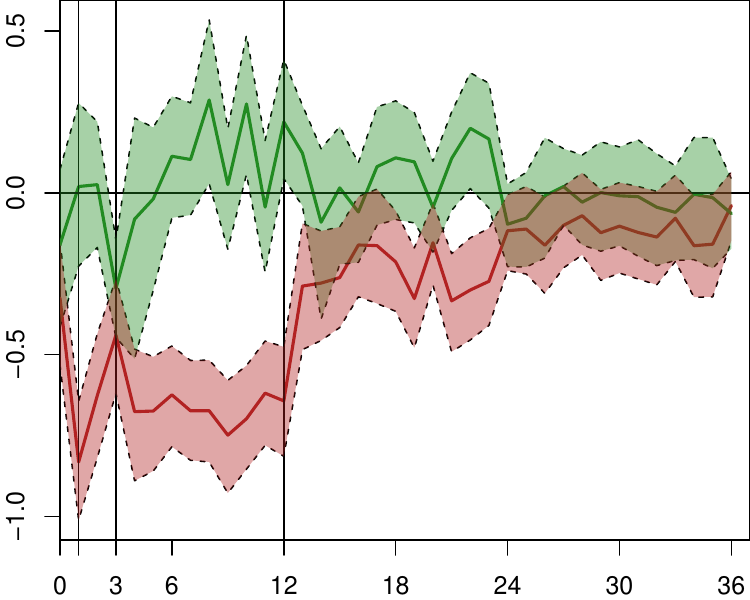}
\end{minipage}
\begin{minipage}{0.49\textwidth}
\centering
\includegraphics[scale=.55]{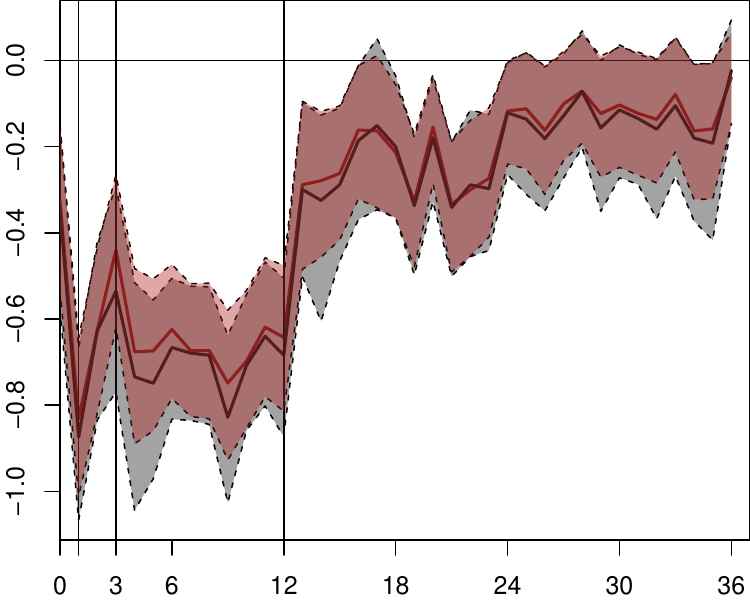}
\end{minipage}
\begin{minipage}{1\textwidth}
\centering
\vspace*{10pt}
\small \textit{Industrial production}
\vspace*{10pt}
\end{minipage}
\begin{minipage}{0.49\textwidth}
\centering
\includegraphics[scale=.55]{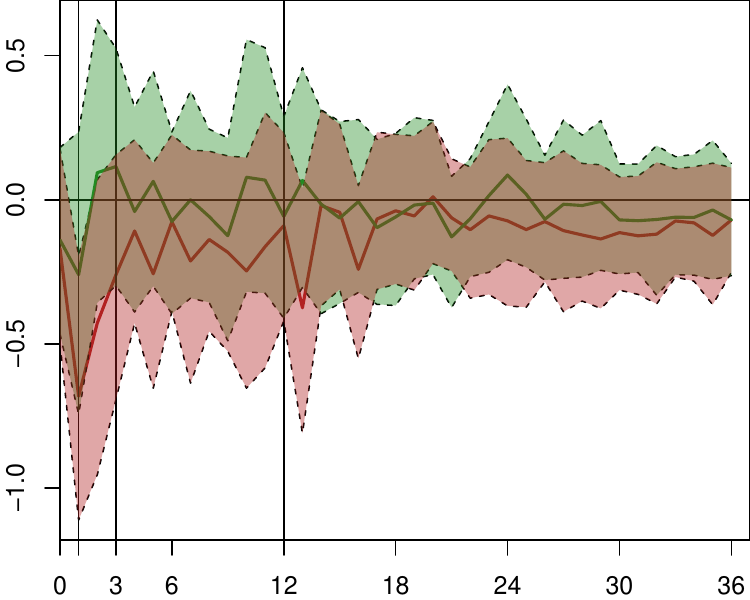}
\end{minipage}
\begin{minipage}{0.49\textwidth}
\centering
\includegraphics[scale=.55]{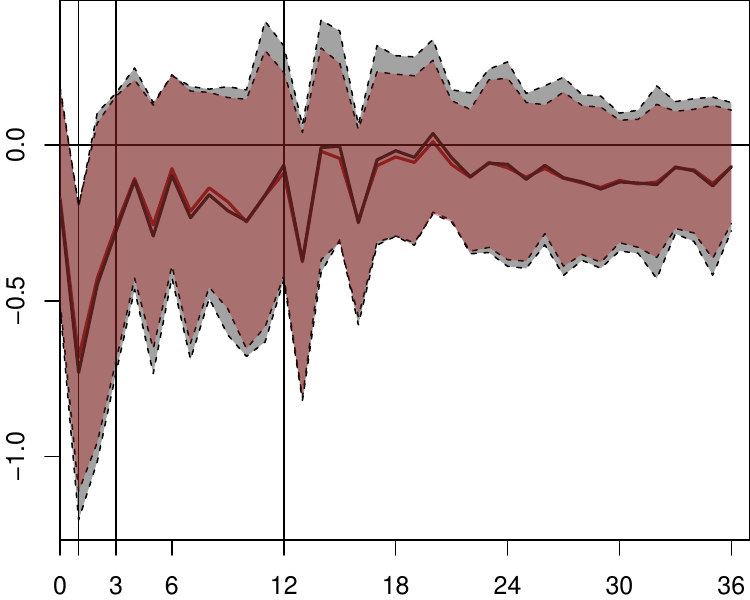}
\end{minipage}
\begin{minipage}{1\textwidth}
\centering
\vspace*{10pt}
\small \textit{Employment}
\vspace*{10pt}
\end{minipage}
\begin{minipage}{0.49\textwidth}
\centering
\includegraphics[scale=.55]{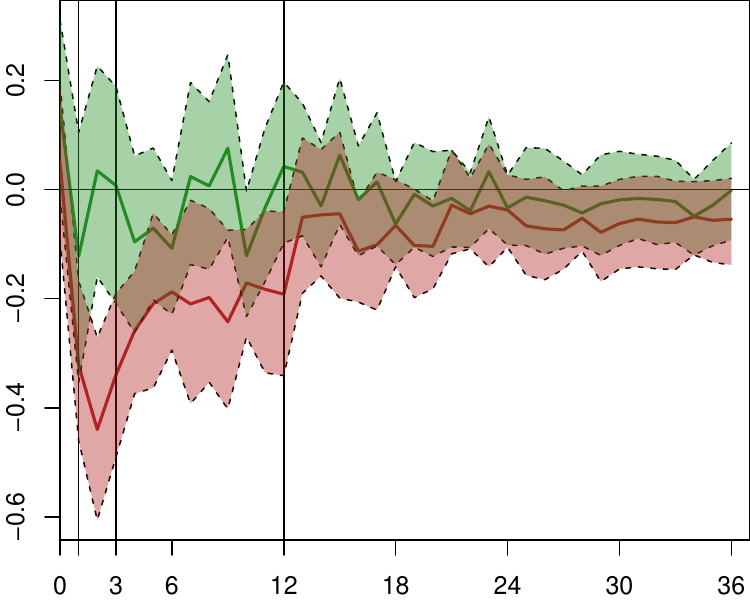}
\end{minipage}
\begin{minipage}{0.49\textwidth}
\centering
\includegraphics[scale=.55]{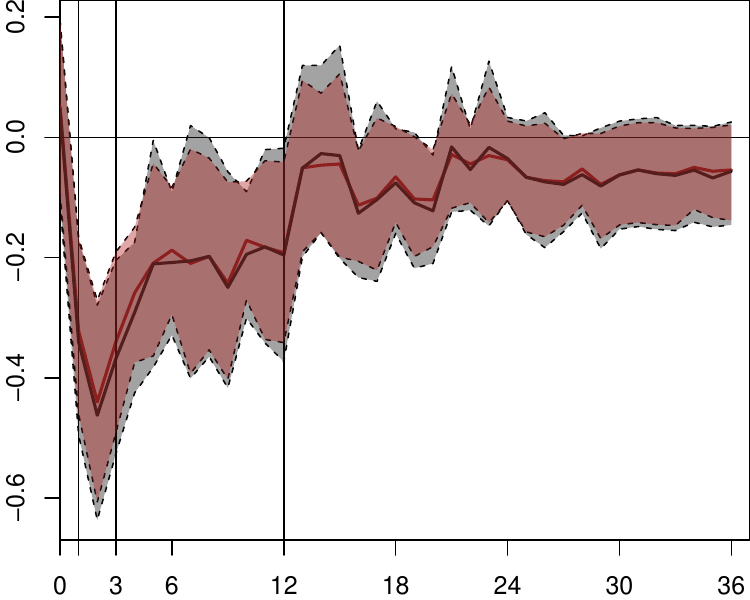}
\end{minipage}
\begin{minipage}{\textwidth}
\vspace{5pt}
\scriptsize \emph{Note:} This figure shows asymmetries in local projection responses in the excessive bond premium (EBP). The left-hand panel refers to asymmetries in the sign of the shock and the right-hand panel refers to asymmetries in the size of the shock. The solid lines indicate the posterior median, while the shaded areas refer to the $68\%$ posterior credible interval.
A positive one-unit shock is denoted in red, a negative one-unit shock is shown in green, and a positive three-unit shock is shown in gray. To ease comparability, we re-scale the responses associated with a negative shock and the responses associated with three-unit shock such that the represent a positive one-unit shock (multiplied by $-1$ and $1/3$, respectively). 
\end{minipage}
\end{figure}

Figure \ref{fig:shlwBNN_lp} shows the 16$^{th}$ and 84$^{th}$ percentiles of the posterior distribution of the LPs (shaded areas) and the median (solid lines)  for different shock signs and shock sizes. Panel (a) shows the responses of inflation, IP and employment to a one unit contractionary financial shock (in red) and to a benign financial shock (in green). The benign shock is multiplied by $-1$ to simplify comparison. Panel (b) shows the responses to a contractionary financial shock to a one and three unit financial shock. The three unit shock is re-scaled by $1/3$ to ease comparison.

The figure reveals substantial asymmetries with respect to the sign of the shock. Consistent with the literature \citep[see, e.g., ][]{barnichon2016theory,barnichon2022effects,forni2024nonlinear}, we find that contractionary shocks have much stronger effects on the macro aggregates than benign shocks. Starting with inflation reactions we find that contractionary financial shocks exert downward pressure on prices. This effect is strong and peaks after around one month, with a  peak median decline in inflation of around 0.8 percentage points. Notice that the reaction is also quite persistent and turns insignificant only after around two years. By contrast, a benign shock leads to a much weaker reaction of inflation. After around one quarter, there is some evidence that inflation picks up. But apart from this, the credible intervals include zero throughout. 

We now turn to the reaction of IP growth. Here, we find much more transient reactions that become insignificant after around a quarter. While we find a negative median reaction of IP growth of around 0.7 percentage points, the reaction to a benign shock are much more muted and never significant.

Similar to the inflation reaction, we find that employment  growth strongly declines after a negative financial shock. This decline peters out after around one year, turning insignificant afterwards. The sharp drop in prices can be linked to the decline in employment and the associated downward pressure on wages. Again, we find no discernible reaction to a benign shock.

In terms of size asymmetries (panel (b) of Figure \ref{fig:shlwBNN_lp}), we find that contractionary financial shocks of different sizes (one and three units) trigger proportional reactions of all three focus variables under consideration.

\section{Conclusion}
Neural networks are extremely popular in many different fields. In economics and econometrics, their usage is growing but still comparably limited. In this paper, we develop techniques to specify and estimate neural networks without requiring cross-validation or much input from the researcher. The key ingredient is a flexible mixture specification on the activation function, which frees the researcher from the necessity to pick a particular activation function. As an additional technical improvement, we allow for heteroskedasticity in the shocks. From a computational point of view, we develop an efficient and scalable (Bayesian) estimation algorithm.

We illustrate how these models can be used for informing policy decisions. In particular, we start by investigating the trade-off between the number of neurons and the number of hidden layers using synthetic data from a realistic DGP that resembles actual US inflation dynamics. We show that, as opposed to theoretical recommendations, a shallow neural network produces favorable MSE ratios and is capable of improving upon deep BNNs.  We then move on to use the models to produce density forecasts of key US macroeconomic aggregates. In this exercise, we find that our BNNs improve upon models commonly used in machine learning, and shallow BNNs perform as well as deep BNNs, while the latter produce much better fit.  The final part of the paper then deals with how BNNs can be used to analyze the effects of economic shocks. To this end, we develop BNN-based local projections and investigate how financial shocks impact US inflation, industrial production and employment. We find substantial asymmetries with respect to sign, with negative shocks exerting stronger effects on inflation and employment than positive ones, but substantial proportionality with respect to the size of the shock.

\clearpage
\small{\setstretch{0.85}
\addcontentsline{toc}{section}{References}
\bibliographystyle{cit_econometrica.bst}
\bibliography{lit}}

\newpage \normalsize
\setcounter{page}{1}
\begin{appendices}
\begin{center}
\LARGE\textbf{Online Appendix}\\[1em]
\Large\textbf{Bayesian Neural Networks for \\ \vspace{-0.3em} Macroeconomic Analysis}\\[1em]
\large\MakeUppercase{Niko Hauzenberger}$^{1}$, \MakeUppercase{Florian Huber}$^2$, \MakeUppercase{Karin Klieber}$^3$, and \MakeUppercase{Massimiliano Marcellino}$^4$\\
$^1$\textit{University of Strathclyde}\\
\vspace{-0.3em} $^2$\textit{University of Salzburg} \\
\vspace{-0.3em} $^3$\textit{Oesterreichische Nationalbank} \\
\vspace{-0.3em} $^4$ \textit{Bocconi University, IGIER, CEPR, Baffi-Carefin and BIDSA}
\end{center}

\setcounter{equation}{0}
\setcounter{table}{0}
\setcounter{figure}{0}
\renewcommand\theequation{A.\arabic{equation}}
\renewcommand\thetable{A.\arabic{table}}
\renewcommand\thefigure{A.\arabic{figure}}

\section{Full conditional posterior distributions}\label{sec: fullpost}
In the following, we provide details on the full conditional posterior distribution for the proposed Markov chain Monte Carlo (MCMC) algorithm outlined in Sub-section \ref{sec: posterior}. 

\begin{itemize}

\item Let $\bm \theta = (\bm \gamma', \bm W_{L+1})'$ denote a $M (= K+Q_L) \times 1$ vector of parameters and $\tilde{\bm x} = (\tilde{\bm x}_1', \dots,\tilde{\bm x}_T')'$ a $T \times M$ matrix of regressors with element $\tilde{\bm x}_t = (\bm x_t', \hat{\bm x}'_t)'$, where 
\begin{equation*}
\hat{\bm x}_t' = \bm h_L\left( \bm W_L \bm h_{L-1} (\cdots \bm W_2 \bm h_1 (\bm W_1 \bm x_t))\right).    
\end{equation*}
Moreover, we define $\bm \Sigma = \text{diag}(\sigma^2_1, \dots, \sigma^2_T)$ as a $T \times T$ matrix capturing the variances and $\underline{\bm V}_{\bm \theta} = \text{diag}(\bm \phi_{\bm \gamma}, \bm \phi_{{L+1}})$ where  $\bm \phi_{\bm \gamma} = (\phi_{\gamma_1},\dots, \phi_{\gamma_K})'$ denotes the $K$ prior variances for the constant coefficients and we collect $\bm \phi_{{L+1}} = (\phi_{L+1,1},\dots, \phi_{L+1,Q_L})'$ for the $Q_L$ nonlinear coefficients. The joint parameter vector $\bm \theta$ is then obtained from a standard multivariate Gaussian posterior: 
\begin{equation}\label{eq:postbeta}
    \bm \theta  | \bullet \sim \mathcal{N}\left(\overline{\bm \theta}, \overline{\bm V}_{\bm \theta}\right),
\end{equation}
with 
\begin{align*}
    \overline{\bm V}_{\bm \theta} &= \left(\tilde{\bm x}' \bm \Sigma^{-1} \tilde{\bm x} + \underline{\bm V}_{\bm \theta}^{-1}\right)^{-1},\\
    \overline{\bm \theta} &= \overline{\bm V}_{\bm \theta} \tilde{\bm x}' \bm \Sigma^{-1} \bm y.
\end{align*}
    
\item The prior on $\bm \gamma$ is of the form:
\begin{equation}\label{eq:postcons}
    \gamma_j \sim \mathcal{N}(0,\phi_{\gamma_j}), \quad \phi_{\gamma_j}  = \lambda^2_{\bm \gamma} \varphi^2_{\gamma_j}, \quad \text{for } j = 1, \dots, M. 
\end{equation}
Similar to $\bm W_{L+1}$, we use a horseshoe prior on $\bm \gamma$ and rely on the hierarchical representation of \cite{makalic2015simple}. The posterior updating step for the hyperparameters $\bm \phi_{\bm \gamma}$ and $\bm \phi_{L+1}$ is essentially the same. Below, we therefore only sketch the updating step for elements in $\bm \phi_{\bm \gamma}$. The global and local shrinkage parameters, $\lambda^2_{\bm \gamma}$ and $\varphi^2_{\gamma_j}$, respectively, are obtained by introducing auxiliary random quantities which follow an inverse Gamma distribution:
\begin{align}
    \varphi^2_{\gamma_j} | \bullet &\sim \mathcal{G}^{-1} \left( 1, c^{-1}_{\gamma_j} + \frac{\gamma^2_{j}}{2 \lambda^2_{\bm \gamma}} \right),\label{eq:postHScons1}\\
    \lambda^2_{\bm \gamma} | \bullet &\sim \mathcal{G}^{-1} \left( \frac{K+1}{2}, d^{-1}_{\bm \gamma} + \sum^K_{j=1} \frac{\gamma^2_{j}}{2  \varphi^2_{\gamma_j}} \right),\label{eq:postHScons2}\\
    c_{\gamma_j} | \bullet &\sim \mathcal{G}^{-1} \left( 1,1+\varphi^{-2}_{\gamma_j}  \right),\label{eq:postHScons3}\\
    d_{\bm \gamma} | \bullet &\sim \mathcal{G}^{-1}\left( 1,1+\lambda^{-2}_{\bm \gamma} \right).\label{eq:postHScons4}
\end{align}

\item To efficiently update $\bm W_{\ell}$ ($\ell = 1, \dots, L)$, we rely on a Hamiltonian Monte Carlo \citep[HMC,][]{neal2011hmc} within Gibbs step.\footnote{\cite{betancourt2018conceptual} provides a general overview of HMC sampling methods.} We therefore iterate through all layers and also take into account the fact that the composite function of $L$ layers has different implications on the gradients of the layer-specific neurons. For each layer-specific $\bm W_{\ell}$, we update the full matrix in a column-wise manner (i.e., $\bm w_{\ell,\bullet j}$, for $j = 1, \dots, Q_{\ell -1}$), which has been found to improve mixing.
In what follows, let $\bm r_{\ell,j}$ denote an auxiliary moment variable for each $\bm w_{\ell,\bullet j}$, where each element is standard normal distributed, i.e., $\bm r_{\ell,j} \sim \mathcal{N}(\bm 0, \bm I)$. Furthermore, let $\mathcal{L}(\bm w_{\ell,\bullet j}) = \log p(\bm w_{\ell,\bullet j} | \bullet)$ define the log conditional posterior density of $\bm w_{\ell,\bullet j}$. 
The fictitious Hamiltonian system of the conditional posterior density of $\bm w_{\ell,\bullet j}$ is then given by 
\begin{equation}\label{eq:hmc}
\mathcal{H}(\bm w_{\ell,\bullet j}, \bm r_{\ell,j}) = -\mathcal{L}(\bm w_{\ell,\bullet j}) + \frac{1}{2} \bm r_{\ell,j}' \bm r_{\ell,j}.
\end{equation}
Since we view Eq. (\ref{eq:hmc}) as a Hamiltonian system, the negative log conditional posterior can be interpreted as a potential energy while the term associated with the moment variable can be interpreted as kinetic energy \citep[see, e.g.,][]{childers2022differentiable}. We simulate the Hamiltonian dynamics of this system via the leapfrog integrator with the following proposals: 
\begin{align}
    \bm r_{\ell,j}^{\ast \ast} &= \bm r_{\ell,j}^{(a)} + \frac{\epsilon}{2} \nabla_{\bm w_{\ell,\bullet j}} \mathcal{L}(\bm w_{\ell,\bullet j}^{(a)}), \\
    \bm w_{\ell,\bullet j}^{\ast} &= \bm w_{\ell,\bullet j}^{(a)} + \epsilon \bm r_{\ell,j}^{\ast \ast}, \\
    \bm r_{\ell,j}^{\ast} &= \bm r_{\ell,j}^{\ast \ast} + \frac{\epsilon}{2} \nabla_{\bm w_{\ell,\bullet j}} \mathcal{L}(\bm w_{\ell,\bullet j}^{\ast}),
\end{align}
where $\bm w_{\ell,\bullet j}^{(a)}$ and $\bm r_{\ell,j}^{(a)}$ denote the previously accepted values, while $\nabla_{\bm w_{\ell,\bullet j}} \mathcal{L}(\bm w_{\ell,\bullet j})$ is the gradient of the log conditional posterior. The leapfrog method uses a discrete step size $\epsilon$ to generate a full-step proposal for $\bm w_{\ell,\bullet j}$ (denoted by $\bm w_{\ell,\bullet j}^{\ast}$) and half-step updates for the momentum $\bm r_{\ell,j}$ (denoted by $ \bm r_{\ell,j}^{\ast \ast}$ for the first half-step update and $\bm r_{\ell,j}^{\ast}$ for the final proposal). We repeat the leapfrog method in $n = 1, \dots, N$ steps. 

The HMC thus uses information of the gradient of the log conditional posterior distribution to propose a new $\bm w_{\ell,\bullet j}$ which greatly improves mixing of the Markov chain. Note that the gradient of the log conditional posterior density can be obtained in a straightforward manner, as we consider only a set of activation functions with well-defined gradients for each neuron. The parameters to tune the HMC algorithm are the step size $\epsilon$ and the number of leapfrog steps $N$. We follow state-of-the-art methods and run the No U-Turn Sampler (NUTS) as proposed by \cite{hoffman2014nuts}, which automatically adapts these tuning parameters during sampling.

Finally, we evaluate the proposed and previously accepted values by means of a Metropolis accept/reject step and determine the acceptance probability $\eta_{\ell,j}$ for proposed $\bm w_{\ell,\bullet j}^{\ast}$:
 \begin{equation}\label{eq:MHstep}
    \eta_{\ell,j} = \text{min} \left( 1, \frac{ \text{exp}( \mathcal{L}(\bm w_{\ell,\bullet j}^{\ast}) - \frac{1}{2} \bm r_{\ell,j}^{'\ast} \bm r_{\ell,j}^{\ast})}{\text{exp}( \mathcal{L}(\bm w_{\ell,\bullet j}^{(a)}) - \frac{1}{2} \bm r_{\ell,j}^{'(a)} \bm r_{\ell,j}^{(a)})} \right).
\end{equation}




\item To achieve shrinkage in the neurons we apply a row-wise horseshoe prior on the elements of $\bm w_{\ell,i\bullet}$ (which denotes the $i^{th}$ row of $\bm W_{\ell}$ for $\ell=1,\dots,L$). We follow \cite{makalic2015simple} and define auxiliary random quantities, which are used to obtain the global and local shrinkage parameters, $\lambda^2_{\ell,i}$ and $\varphi^2_{\ell,ij}$. The (hyper)parameters follow an inverse Gamma distribution:
\begin{align}
    \varphi^2_{\ell,ij} | \bullet &\sim \mathcal{G}^{-1} \left( 1, c^{-1}_{\ell,ij} + \frac{w_{\ell,ij}}{2 \lambda^2_{\ell,i}} \right),\label{eq:postHS1}\\
    \lambda^2_{\ell,i} | \bullet &\sim \mathcal{G}^{-1} \left( \frac{Q_\ell+1}{2}, d_{\ell,i}^{-1} + \sum^{Q_\ell}_{j=1} \frac{w^2_{\ell,ij}}{2 \varphi^2_{\ell,ij}} \right),\label{eq:postHS2}\\
    c_{\ell,ij} | \bullet &\sim \mathcal{G}^{-1} \left( 1,1+\varphi^{-2}_{\ell,ij} \right),\label{eq:postHS3}\\
    d_{\ell,i} | \bullet &\sim \mathcal{G}^{-1}\left( 1,1+\lambda^{-2}_{\ell,i} \right).\label{eq:postHS4}
\end{align}

\item For choosing the activation function $h_{\ell,q}$ we draw the indicator $\delta_{\ell,q}$ from a multinomial distribution of the following form:
\begin{equation}\label{eq:postMxInd}
\text{Pr}(\delta_{\ell,q} = m | \bullet) \propto \omega_{\ell, qm} \times \exp \left \{- \frac{1}{2}\left(\bm (\bm y - \hat{\bm y}^{(m)}_{\ell, q})' \bm \Sigma^{-1} (\bm y - \hat{\bm y}^{(m)}_{\ell, q}) \right) \right\}, 
\end{equation}
for $m = 1, \dots, 4$, where $\hat{\bm y}^{(m)}_{\ell,q} = (\hat{y}^{(m)}_{\ell q,1}, \dots, \hat{y}^{(m)}_{\ell q,T})'$ with elements $\hat{y}^{(m)}_{\ell q,t} = w_{\ell,q} h^{(m)}(z_{\ell q,t})$.
\end{itemize}

\setcounter{equation}{0}
\setcounter{table}{0}
\setcounter{figure}{0}
\renewcommand\theequation{B.\arabic{equation}}
\renewcommand\thetable{B.\arabic{table}}
\renewcommand\thefigure{B.\arabic{figure}}

\section{Technical details on the benchmark models}
\subsection{Bayesian neural network by backpropagation (BNN-BP)}\label{sec: BBB}

A Bayesian neural network estimated by backpropagation was introduced by \cite{blundell2015weight}. It serves as a variational inference scheme for learning the posterior distribution on the weights of a neural network. To do so, the approach maximizes the log-likelihood of the model subject to a Kullback-Leibler complexity term on the parameters and makes use of the reparameterization trick \citep{kingma2013reparam} to obtain the posterior distribution of the weights with stochastic gradient descent. 

The prior on the weights is specified as a scale mixture of two Gaussian densities with zero mean but differing variances. The first mixture component features a large variance ($\sigma^2_{BP,1} = 3$) providing a heavy-tailed distribution whereas the variance of the other component is set small ($\sigma^2_{BP,2} = 0.0025$) concentrating the weights a priori around zero. This setup is similar to a spike and slab prior \citep[see,][]{george1993variable} but with the same prior parameters for all the weights to allow for the optimization by stochastic gradient descent. Note that the horseshoe prior, which we use in our proposed BNN model, approximates the spike and slab prior. This implies that the main difference between BNN-BP and our proposed model lies in the inference method, where the former relies on optimization techniques to approximate the posterior, while the latter uses MCMC to generate samples from the posterior distribution directly. 

The hyperparameters are chosen in a cross validation exercise based on an expanding window time series split. Specifically, we use all observations up to the last 24 months before the start of our hold-out to train the model and then, after obtaining the predictive densities, add the next observation and recompute the model. We repeat this until we end up at the beginning of our hold-out and choose the specification with the lowest average RMSE.
We train all models in 1000 epochs and use the MSE loss function, the ADAM optimizer and a learning rate of 0.01.

\subsection{Tree ensemble methods}\label{sec: BART}
Tree ensembles are an alternative to BNNs to approximate the unknown function $f$. In our set of competitors we consider Bayesian additive regression trees \citep[BART, ][]{chipman2010bart} as well as random forests \citep[RF,][]{breiman2001random}. While both BART and RF are ensemble methods, they differ in their modeling approaches, uncertainty estimation and complexity. 

BART is based on a Bayesian framework, providing uncertainty estimates for its predictions through the posterior distribution of model parameters, where regression trees are combined additively. A random forest typically does not offer a formal probabilistic framework for uncertainty estimation. It combines multiple independently estimated regression trees by averaging across the individual predictions. This implies that BART may capture more complex nonlinear relationships due to its additive nature, while RF is often more straightforward to implement and less computationally intensive. For the latter, we apply the R package \texttt{randomForest} of \cite{liaw2002randomforest}.

In the case of BART, the idea is to consider the sum over a number of $Z$ regression trees. Formally, this boils down to:
\begin{equation}
f(\bm x_t) \approx \sum^Z_{z=1} g_z (\bm x_t | \mathcal{T}_{z}
, \bm \rho_{z}) 
\end{equation}
A single regression tree $g_z$ depends on two parameters, the tree structure given by $\mathcal{T}_{z}$ and the terminal node parameter $\bm \rho_{z}$. Following \cite{chipman2010bart}, we set $Z=250$ and build the prior on the tree structure upon a tree-generating stochastic process. This involves determining the probability that a given node is nonterminal, the selection of variables used in a splitting rule (to spawn left and right children nodes) and the corresponding thresholds. For the terminal node parameter we specify a conjugate Gaussian prior distribution with data-based prior variance. In particular, the specification centers prior mass on the range of the data while ensuring a higher degree of shrinkage if the number of trees is large. Details can be found in \cite{chipman2010bart}. 


\setcounter{equation}{0}
\setcounter{table}{0}
\setcounter{figure}{0}
\renewcommand\theequation{C.\arabic{equation}}
\renewcommand\thetable{C.\arabic{table}}
\renewcommand\thefigure{C.\arabic{figure}}


\section{Fluctuation tests}\label{sec: app_fluctuation}

\begin{figure}[!htbp]
\caption{Fluctuation test statistic for BNNs and the best performing benchmark. \label{fig:fluct_test}}

\begin{minipage}{0.32\textwidth}
\centering
\scriptsize \textit{Inflation}
\end{minipage}
\begin{minipage}{0.32\textwidth}
\centering
\scriptsize \textit{Industrial production}
\end{minipage}
\begin{minipage}{0.32\textwidth}
\centering
\scriptsize \textit{Employment}
\end{minipage}

\begin{minipage}{0.32\textwidth}
\centering
\includegraphics[scale=.4]{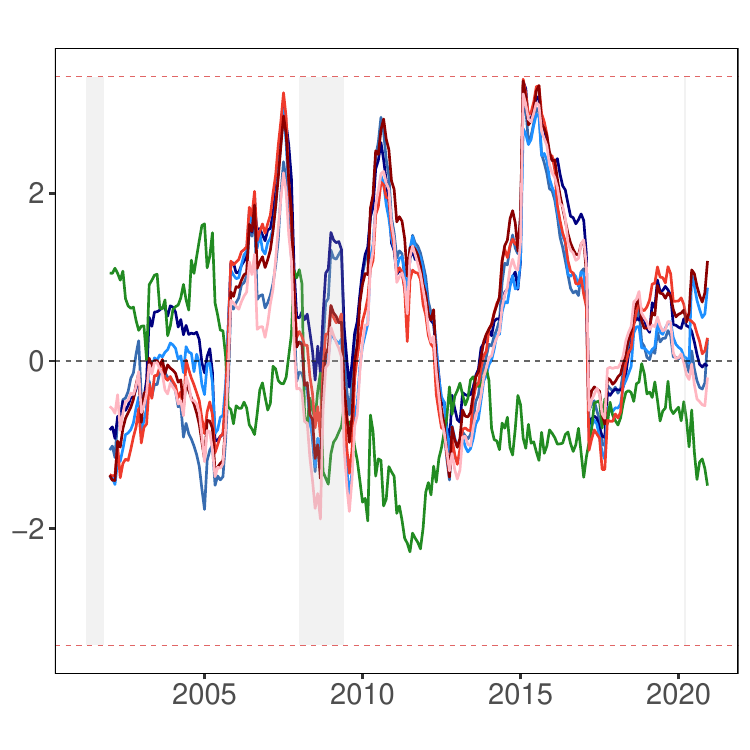}
\end{minipage}
\begin{minipage}{0.32\textwidth}
\centering
\includegraphics[scale=.4]{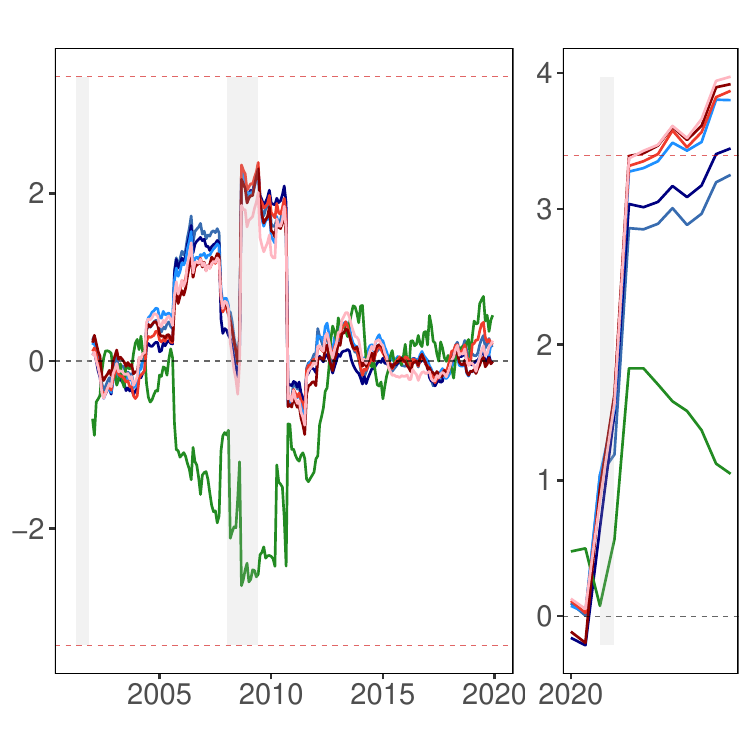}
\end{minipage}
\begin{minipage}{0.32\textwidth}
\centering
\includegraphics[scale=.4]{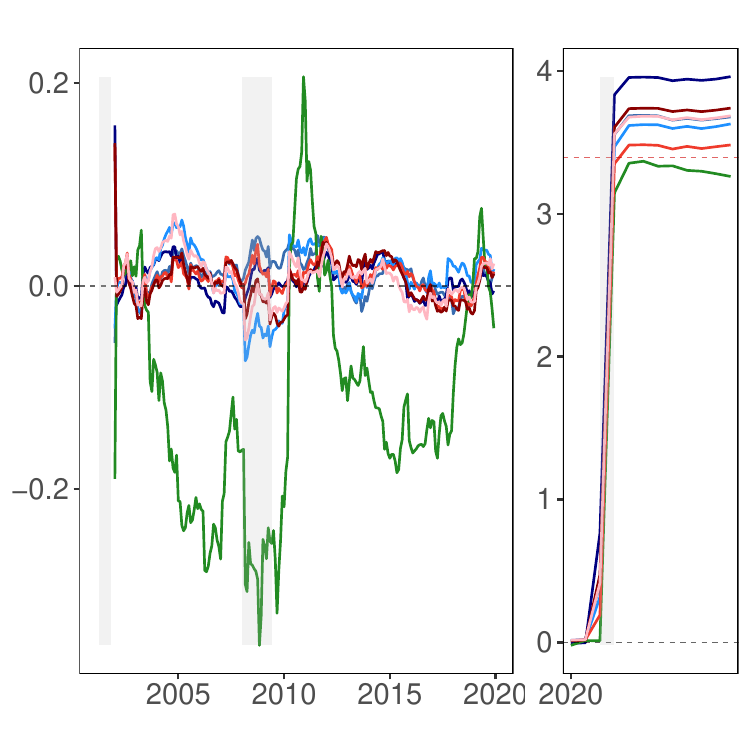}
\end{minipage}

\begin{minipage}{\textwidth}
\centering
\vspace{-0.8cm}
\includegraphics[scale=.4]{LPL_legend_col.pdf}
\end{minipage}

\begin{minipage}{\textwidth}
\vspace{2pt}
\scriptsize \emph{Note:} This figure shows the fluctuation test statistic relative to the benchmark as proposed by \cite{giacomini2010forecast}. Dashed lines indicate critical values for a 5\% level of statistical significance. Positive values of the fluctuation test imply that the corresponding model outperforms the benchmark.
\end{minipage}
\end{figure}

To test whether the outperformance presented in Figure \ref{fig:oos_lps} is stable over time, we calculate the fluctuation test statistic as proposed by \cite{giacomini2010forecast}. Again, the linear model with the horseshoe prior serves as the benchmark in each application. We choose the 5 percent level for statistical significance. Positive values of the fluctuation test statistic imply that the corresponding model outperforms the benchmark. If model performance is inferior the test statistic gives negative values. Results are presented in Figure \ref{fig:fluct_test}.

\end{appendices}
\end{document}